\documentclass[aps,twocolumn,groupedaddress]{revtex4}
\usepackage{graphicx}
\usepackage{amsmath}
\usepackage{amssymb}
\usepackage{amsfonts}

\begin{document}

\title{Implementing quantum information processing with atoms, ions and
photons \footnote{les Houches summer school proceedings, session
79, Ed. D. Est{\`e}ve, J.M. Raimond and J. Dalibard, Elsevier,
Amsterdam (2004) }}
\author{P. Zoller$^{1}$, J. I. Cirac$^{2}$, Luming Duan$^{3}$ and J.~J. Garc%
\'{\i}a-Ripoll$^{2}$  } \affiliation{$^{1}$Institute for
Theoretical Physics, University of Innsbruck, and Institute of the
Austrian Academy of Sciences for Quantum Optics and Quantum
Information, A-6020 Innsbruck, Austria} \affiliation{$^{2}$
Max-Planck-Institut f\"ur Quantenoptik, Hans-Kopfermann-Str. 1,
Garching, D-85748, Germany } \affiliation{$^{3}$Department of
Physics and FOCUS center, University of Michigan, Ann Arbor, MI
48109}
\date{2003 08}
\maketitle


\section{Introduction}

Quantum optical systems are one of the very few examples of quantum systems,
where complete control on the single quantum level can be realized in the
laboratory, while at the same time avoiding unwanted interactions with the
environment causing decoherence. These achievements are illustrated by
storage and laser cooling of single trapped ions and atoms, and the
manipulation of single photons in Cavity QED, opening the field of
engineering interesting and useful quantum states. In the mean time the
frontier has moved towards building larger composite systems of a few atoms
and photons, while still allowing complete quantum control of the individual
particles. The new physics to be studied in these systems is based on
entangled states, both from a fundamental point of testing quantum mechanics
for larger and larger systems, but also in the light of possible new
applications like quantum information processing or precision measurements%
\cite{Nielsen+Chuang-QC:00,Braunstein+Pati-QuantInfoContVar:03}.

Guided by theoretical proposals as reviewed in \cite%
{Cirac+Duan+Jaksch+Zoller-Varenna:02}, we have seen extraordinary progress
in experimental AMO physics during the last few years in implementing
quantum information processing. Highlights are the recent accomplishments
with ion traps\cite{Levi-PhysTodayIonTrapQC:03}, cold atoms in optical
lattices \cite{Cirac+Zoller-ScienceColdAtomReview:03}, cavity QED (CQED)
\cite{Raimond+Brune+Haroche-collRMP:01} and atomic ensembles \cite%
{Lukin-AtomicEnsemblesCollRevModPhys:03}. Below we summarize some of the
theoretical aspects of implementing quantum information processing with
quantum optical systems. In particular, in Sec. \ref{sec:ions} we discuss
quantum computing with trapped ions. Sec. \ref{sec:coldatoms} demonstrates
cold coherent collisions as a mean to entangle atoms in an optical lattice.
Finally, Sec. \ref{sec:atomicensembles} reviews atomic ensembles.

\section{Trapped Ions}

\label{sec:ions}

Trapped ions is one of the most promising systems to implement quantum
computation \cite{Cirac+Zoller-QuantCompColdTrappedIons:95,
Steane-ReviewIonTrap:97, Cirac+Duan+Jaksch+Zoller-Varenna:02}. In this
section we describe the theory of quantum information processing with a
system of trapped ions. On the experimental side remarkable progress has
been reported during the last two years in realizing some of the these ideas
in the laboratory\cite{Schmidt-Kaler+BlattETAL-CiracZollerGate:03,
Leibfried+DeMarcoETAL-ExpDemTwoIonPhaseGate:03,
Leibfried+DeMarcoETAL-NISTreview:03}, as explained in the lecture notes by
R. Blatt and D. Wineland.

Ion trap quantum computing, as first proposed in Ref.~ \cite%
{Cirac+Zoller-QuantCompColdTrappedIons:95}, stores qubits in longlived
internal states of single trapped ion. Single qubit gates are performed by
coupling the qubit states to laser light for an appropriate period of time.
In general, this requires that single ions can be addressed by laser light.
Two qubit gates can be achieved by entangling ions via collective phonon
modes. Depending on the specific protocol this requires the initialization
of the phonon bus in a pure initial state, e.g. laser cooling to the
motional ground state in ion traps. However, recently specific protocols for
\textquotedblleft hot gates\textquotedblright\ have been developed which
loosen these requirements (see \cite{Cirac+Duan+Jaksch+Zoller-Varenna:02,
Garcia-Ripoll+Zoller+Cirac-SpeedOptimizedIonGate:03} and references cited).
The unitary operations, which can be decomposed in a series of single and
two-qubit operations on the qubits, can either be performed \emph{dynamically%
}, i.e. based on the time evolution generated by a specific Hamiltonian, or
\emph{geometrically} as in holonomic quantum computing \cite{Duan01}.
Finally, read out of the atomic qubit is accomplished using the method of
quantum jumps \cite{Gardiner+Zoller-QuantumNoise:99}.

An essential feature of ion trap quantum computers is the scalability to a
large number of qubits. This is achieved by moving ions from a storage area,
either to address the ions individually to perform the single qubit
rotation, or by bringing pairs of ions together to perform a two-qubit gate.
Moving ions does not affect the qubit stored in the internal electronic or
hyperfine states, and heating of the ion motion can cooled in a
nondestructive way by sympathetic cooling \cite%
{Wineland+MonroeETAL-NISTbible:98, Cirac+Zoller-ScalableQCTrappedIons:00,
Kielpinksi+Monroe+Wineland-ScalableQC:02}.

In our discussion below we will start with a brief outline of manipulation
of trapped ions by laser light. We then proceed to illustrate ion trap
quantum computing with two specific examples. We will first discuss in some
detail the basic physical ideas and requirements of the original ion trap
proposal\cite{Cirac+Zoller-QuantCompColdTrappedIons:95}. Our emphasis is on
the two qubit gate, and in direct relation to experimental work described by
R. Blatt and D. Wineland. As a second example, we discuss the most recent
proposal for a fast and robust 2-qubit gates for scalable ion trap quantum
computing, based on laser coherent control techniques \cite%
{Garcia-Ripoll+Zoller+Cirac-SpeedOptimizedIonGate:03}. This 2-qubit gate can
be orders of magnitude faster than the time scale given by the trap period,
thus overcoming previous speed limits of ion trap quantum computing, while
at the same time relaxing the experimental constraints of individual laser
addressing of the ions and cooling to low temperatures.

\subsection{Modelling a single trapped ion}

In this section we give a theoretical description of quantum state
engineering in a system of trapped and laser cooled ions. The development of
the theory begins with the description of Hamiltonians, state preparation,
laser cooling and state measurements for single ions, and then followed by a
generalization to the case of many ions. This serves as the basis of our
discussion of quantum computer models.

We describe a single trapped ion driven by laser light as a two-level atom $%
|g\rangle,|e\rangle$ moving in a 1D harmonic confining potential \cite%
{Cirac+Duan+Jaksch+Zoller-Varenna:02} with Hamiltonian ($\hbar=1$)
\begin{equation}
H=\nu a^{\dagger}a-\frac{1}{2}\Delta\sigma_{z}+\frac{1}{2}\Omega\{\sigma
_{+}e^{i\eta(a+a^{\dagger})}+\text{h.c.}\}.   \label{Hsingleion}
\end{equation}
Here the first term is the harmonic oscillator Hamiltonian for the
center-of-mass motion of the ion with trap frequency $\nu$. We have denoted
by $a$ and $a^{\dag}$ the lowering and raising operators, respectively,
which can be expressed in terms of the position and momentum operators as $%
\hat{x}=\sqrt{1/2M\nu}(a+a^{\dagger})$ and $\hat{p}=i\sqrt{M\nu/2}%
(a^{\dagger}-a)$ with $M$ the ion mass. The second and third term in \ref%
{Hsingleion} describe the driven two-level system in a rotating frame using
standard spin-$\frac {1}{2}$ notation, $\sigma_{+}=(\sigma_{-})^{\dagger}=|e%
\rangle\langle g|$ and $\sigma_{z}=|e\rangle\langle e|-|g\rangle\langle g|$.
This internal atomic Hamiltonian is written in a frame rotating with the
optical frequency. We denote by $\Delta=\omega_{L}-\omega_{eg}$ the detuning
of the laser with $\omega_{L}$ the laser frequency and by $\omega_{eg}$ the
atomic transition frequency, and $\Omega$ is the Rabi frequency for the
transition $|g\rangle\rightarrow|e\rangle$. In writing \ref{Hsingleion} we
have assumed that the atom is driven by a running laser wave with wave
vector $k_{L}=2\pi/\lambda_{L}$ along the oscillator axis. Transitions from $%
|g\rangle$ to $|e\rangle$ are associated with a momentum kick to the atom by
absorption of a laser photon, as described by $\exp(ik_{L}\hat{x}%
)\equiv\exp(i\eta (a+a^{\dagger}))$, which couples the motion of the ion
(phonons) to the internal laser driven dynamics.

\begin{figure}[pbh]
\label{ion1}
\par
\begin{center}
\includegraphics[width=8.0cm]{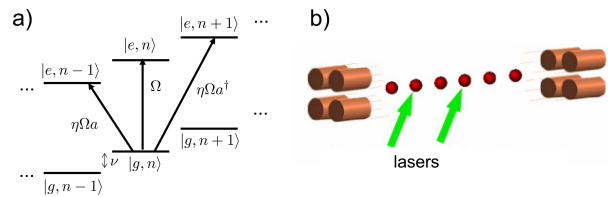}
\end{center}
\caption{a) Coupling to the atom + trap levels according to the Hamiltonians
(\protect\ref{h0}), (\protect\ref{kone} and (\protect\ref{kmone},
respectively, in lowest order Lamb-Dicke expansion. b) ion trap quantum
computer 1995 (schematic)}
\end{figure}

In Eq.~(\ref{Hsingleion}) we have defined a Lamb-Dicke parameter $\eta= 2
\pi a_{0} / \lambda_{L}$ with $a_{0} 0 \sqrt{1/2M\nu}$ the ground state size
of the oscillator and $\lambda_{L}$ the laser wave length. In the Lamb-Dicke
limit $\eta\ll1$ we can expand the atom laser interaction: $H_{AL}=\frac{1}{2%
}\Omega\{\sigma_{-}[1+i\eta(a+a^{\dagger})+\mathcal{O}(\eta^{2})]+$h.c.$\}.$
The resulting Hamiltonian can be further simplified if the laser field is
sufficiently weak so that only pairs of bare atom + trap levels are coupled
resonantly. We denote by $|g\rangle|n\rangle$ and $|e\rangle|n\rangle$ the
eigenstates of the bare Hamiltonian $H_{0}=\nu a^{\dagger}a-\frac{1}{2}%
\Delta\sigma_{z}$, where the internal two-level system is in the ground
(excited) state and $n$ is the phonon excitation number of the harmonic
oscillator. When tuning the laser to atomic resonance $\Delta=\omega
_{L}-\omega_{eg}\approx0$, i.e. $|\omega_{L}-\omega_{eg}|\ll\nu$, the
transitions changing the harmonic oscillator quantum number $n$ are
off-resonance and can be neglected. In this case the Hamiltonian (\ref%
{Hsingleion}) can be approximated by
\begin{equation}
H_{0}=\nu a^{\dagger}a-\frac{1}{2}\Delta\sigma_{z}+\frac{1}{2}\Omega
(\sigma_{+}+\text{\textrm{h.c.}})\quad(\Delta\approx0),   \label{h0}
\end{equation}
On the other hand, for laser frequencies close to the lower (red) motional
sideband resonance $\Delta\approx-\nu$, i.e. $|\omega_{L}-(\omega_{eg}-\nu)|%
\ll\nu$, only transitions decreasing the quantum number $n$ by one are
important, and $H$ can be approximated by a Hamiltonian of the
Jaynes-Cummings type:
\begin{equation}
H_{\text{JC}}{}=\nu a^{\dag}a-\frac{1}{2}\Delta\sigma_{z}+\frac{1}{2}%
\Omega(i\eta\sigma_{+}a+\text{\textrm{h.c.}})\quad(\Delta\approx -\nu).
\label{kone}
\end{equation}
Similarly, for tuning to the upper (blue) sideband $\Delta\approx+\nu$ $,$%
i.e. $\omega_{L}-(\omega_{eg}+\nu)|\ll\nu$, only transitions increasing the
quantum number $n$ by one contribute, so that $H$ can be approximated by the
\emph{anti}-Jaynes-Cummings Hamiltonian
\begin{equation}
H_{\text{AJC}}{}=\nu a^{\dagger}a-\frac{1}{2}\Delta\sigma_{z}+\frac{1}{2}%
\Omega(i\eta\sigma_{+}a^{\dagger}+\text{\textrm{h.c.}})\quad(\Delta
\approx+\nu).   \label{kmone}
\end{equation}
(see Fig.~\ref{ion1}) For the above approximations to be valid we require
that the effective Rabi frequencies to the non-resonant states have to be
much smaller than the trap frequency, i.e. we must spectroscopically resolve
the motional sidebands.

The eigenstates of the Hamiltonians $H_{0}$, $H_{\text{JC}}{}_{\pm}$ and $H_{%
\text{AJC}}{}_{\pm}$ are the dressed states. These states are familiar from
cavity QED, and are obtained by diagonalizing the $2x2$ matrices of nearly
degenerate states. Applying a laser pulse on resonance, $\Delta=0$, will
according to (\ref{h0}) induce Rabi flopping between the states $%
|g\rangle|n\rangle$ and $|e\rangle|n\rangle$, while a laser tuned for
example to the lower motional sideband $\Delta=-\nu$ will lead to Rabi
oscillations coupling $|g\rangle|n\rangle$ and $|e\rangle|n-1\rangle$. The
above Hamiltonians are basic building blocks to engineer general quantum
states of motion. As an example, a laser pulse applied on the carrier
frequency ($\Delta=0$) to a state $(\alpha|g\rangle+\beta|e\rangle)|0\rangle$
will induce a general Rabi rotation without affecting the phonon state, i.e.
perform a single qubit rotation. On the other hand, a $\pi$-pulse with
duration $T=\pi/\eta\Omega$ on the red sideband will swap an initial
superposition of qubits to a corresponding superposition of phonon states, $%
(\alpha|g\rangle+\beta|e\rangle)|0\rangle\rightarrow|g\rangle(\alpha
|0\rangle+\beta|1\rangle)$. These processes will be the building blocks for
the quantum gate discussed below.

We note that the interaction time for the above processes must always be
much longer than the trap period $1/\nu$. On the other hand, when we apply a
short laser pulse to the ion much less than the trap period $1/\nu$, i.e. we
do not spectroscopically resolve the sidebands, and we can ignore the trap
motion during the time duration of the pulse. A $\pi$-pulse to the two level
atom is thus accompanied by a momentum kick to the motional state $|g\rangle
|$motion$\rangle\rightarrow|e\rangle e^{ik_{L}\hat{x}}|$motion$\rangle$, $%
|e\rangle|$motion$\rangle\rightarrow|e\rangle e^{-ik_{L}\hat{x}}|$motion$%
\rangle$. In particular, if we choose a coherent state $|\alpha \rangle_{%
\text{coh}}$ to represent the motion, we will shift the coherent states $%
|g\rangle|\alpha\rangle_{\text{coh}}\rightarrow|e\rangle|\alpha
+i\eta\rangle_{\text{coh}}$, $|e\rangle|\alpha\rangle_{\text{coh}%
}\rightarrow|g\rangle|\alpha-i\eta\rangle_{\text{coh}}$. Furthermore, if we
apply a short $\pi$-pulse in the direction $+k_{L}$ followed by a pulse from
the opposite direction $-k_{L}$, we achieve a transformation
\begin{align*}
|g\rangle|\alpha\rangle_{\text{coh}} & \rightarrow|g\rangle|\alpha
+2i\eta\rangle_{\text{coh}} \\
|e\rangle|\alpha\rangle_{\text{coh}} & \rightarrow|e\rangle|\alpha
-2i\eta\rangle_{\text{coh}}
\end{align*}
This process will be the basic element of the high speed 2 qubit gate at the
end of this section.

\subsection{Ion trap quantum computer '95}

\begin{figure}[pbh]
\label{ion2}
\par
\begin{center}
\includegraphics[width=8.0cm]{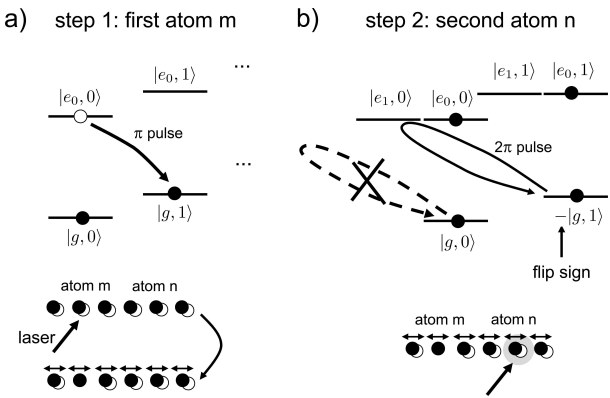}
\end{center}
\caption{The two-qubit quantum gate with trapped ions \protect\cite%
{Cirac+Zoller-QuantCompColdTrappedIons:95}. a) First step according to(%
\protect\ref{bigone}): the qubit of the first atom is swapped to the
photonic data bus with a $\protect\pi$-pulse on the lower motional sideband,
b) Second step: the state $|g,1\rangle$ acquires a minus sign due to a $2%
\protect\pi$-rotation via the auxiliary atomic level $|r_{1}\rangle$ on the
lower motional sideband.}
\end{figure}

We describe in some detail the 2-qubit gate in the original ion trap
proposal, as illustrated in Fig.~\ref{ion2}. \cite%
{Cirac+Zoller-QuantCompColdTrappedIons:95}. In the ion trap quantum
computer'95 qubits are represented by the long-lived internal states of the
ions, with $|g\rangle_{j}\equiv|0\rangle_{j}$ the ground state, and $%
|e_{0}\rangle_{j}\equiv|1\rangle_{j}$ a (metastable) excited state ($%
j=1,\ldots,N$). In addition, we assume that there is a second metastable
excited state $|e_{1}\rangle$ which plays below the role of an auxiliary
state. In this system separate manipulation of each individual qubit is
accomplished by addressing the ions with different laser beams and inducing
a Rabi rotation. The heart of the proposal is the implementation of a
two-qubit gate between two (or more) arbitrary ions in the trap by exciting
the collective quantized motion of the ions with lasers, i.e. the collective
phonon mode plays the role of a quantum data bus. For this we assume that
the collective phonon modes have been cooled initially to the ground state.

Single qubit rotations can be performed tuning a laser on resonance with the
internal transition ($\Delta_{j}=0$) with polarization $q=0$, $|g\rangle
_{j}\rightarrow|e_{0}\rangle_{j}$. In an interaction picture the
corresponding Hamiltonian is
\begin{equation}
\hat{H}_{j}=\frac{1}{2}\Omega\left[ |e_{0}\rangle_{j}\langle g|e^{-i\phi
}+|g\rangle_{j}\langle e_{0}|e^{i\phi}\right] .   \label{Ha}
\end{equation}
For an interaction time $t=k\pi/\Omega$ (i.e., using a $k\pi$ pulse), this
process is described by the following unitary evolution operator
\begin{equation}
\hat{V}_{j}^{k}(\phi)=\exp\left[ -ik\frac{\pi}{2}(|e_{0}\rangle_{j}\langle
g|e^{-i\phi}+h.c.)\right] \;,   \label{Vn}
\end{equation}
so that we achieve a Rabi rotation
\begin{align}
|g\rangle_{j} & \longrightarrow|g\rangle_{j}\cos(k\pi/2)-|e_{0}\rangle
_{j}ie^{i\phi}\sin(k\pi/2),  \notag \\
|e_{0}\rangle_{j} & \longrightarrow|e_{0}\rangle_{j}\cos(k\pi/2)-|g\rangle
_{j}ie^{-i\phi}\sin(k\pi/2).  \notag
\end{align}

When we work with $N$ ions, the ion chain supports $N$ longitudinal modes,
of which the center of mass mode, $\nu_{1}=\nu$, is energetically separated
from the rest, $\nu_{k}>=\sqrt{3}\nu$ ($k>1$). If the laser addressing the $j
$-th ion is tuned to the lower motional sideband of, for example, the
center-of-mass mode, we have in the interaction picture the Hamiltonian
\begin{equation}
{H}_{j,q}=\frac{\eta}{\sqrt{N}}\frac{\Omega}{2}\left[ |e_{q}\rangle
_{j}\langle g|ae^{-i\phi}+|g\rangle_{j}\langle e_{q}|a^{\dagger}e^{i\phi }%
\right] .   \label{Hb}
\end{equation}
Here $a^{\dagger}$ and $a$ are the creation and annihilation operator of the
center-of-mass phonons, respectively, $\Omega$ is the Rabi frequency, $\phi$
the laser phase, and $\eta$ is the Lamb-Dicke parameter. The subscript $%
q=0,1 $ refers to the transition excited by the laser, which depends on the
laser polarization.

If this laser beam is on for the time $t=k\pi/(\Omega\eta/\sqrt{N})$ (i.e.,
using a $k\pi$ pulse), the evolution of the system will be described by the
unitary operator:
\begin{equation}
\hat{U}_{j}^{k,q}(\phi)=\exp\left[ -ik\frac{\pi}{2}(|e_{q}\rangle_{j}\langle
g|ae^{-i\phi}+h.c.)\right] .   \label{Un}
\end{equation}
It is easy to prove that this transformation keeps the state $|g\rangle
_{j}|0\rangle$ unaltered, whereas
\begin{align}
|g\rangle_{j}|1\rangle & \longrightarrow|g\rangle_{j}|1\rangle\cos
(k\pi/2)-|e_{q}\rangle_{j}|0\rangle ie^{i\phi}\sin(k\pi/2),  \notag \\
|e\rangle_{j}|0\rangle & \longrightarrow|e_{q}\rangle_{j}|0\rangle\cos
(k\pi/2)-|g\rangle_{j}|1\rangle ie^{-i\phi}\sin(k\pi/2),  \notag
\end{align}
where $|0\rangle$ ($|1\rangle$) denotes a state of the CM mode with no (one)
phonon.

Let us now show how a two-bit gate can be performed using this interaction.
We consider the following three--step process (see Fig.~\ref{ion2}):

\begin{description}
\item[(i)] A $\pi$ laser pulse with polarization $q=0$ and $\phi=0$ excites
the $m$-th ion. The evolution corresponding to this step is given by $\hat
{%
U}_{m}^{1,0}\equiv\hat{U}_{m}^{1,0}(0)$ (Fig.~\ref{ion2}a).

\item[(ii)] The laser directed on the $n$--th ion is then turned on for a
time of a $2\pi$-pulse with polarization $q=1$ and $\phi=0$. The
corresponding evolution operator $\hat{U}_{n}^{2,1}$ changes the sign of the
state $|g\rangle_{n}|1\rangle$ (without affecting the others) via a rotation
through the auxiliary state $|e_{1}\rangle_{n}|0\rangle$ (Fig.~\ref{ion2}b).

\item[(iii)] Same as (i).
\end{description}

Thus, the unitary operation for the whole process is $\hat{U}_{m,n}\equiv
\hat{U}_{m}^{1,0}\hat{U}_{n}^{2,1}\hat{U}_{m}^{1,0}$ which is represented
diagrammatically as follows:
\begin{equation*}
\begin{array}[b]{rrrrr}
& \hat{U}_{m}^{1,0} &  & \hat{U}_{n}^{2,1} &  \\
|g\rangle _{m}|g\rangle _{n}|0\rangle  & \longrightarrow  & |g\rangle
_{m}|g\rangle _{n}|0\rangle  & \longrightarrow  & |g\rangle _{m}|g\rangle
_{n}|0\rangle  \\
|g\rangle _{m}|e_{0}\rangle _{n}|0\rangle  & \longrightarrow  & |g\rangle
_{m}|e_{0}\rangle _{n}|0\rangle  & \longrightarrow  & |g\rangle
_{m}|e_{0}\rangle _{n}|0\rangle  \\
|r_{0}\rangle _{m}|g\rangle _{n}|0\rangle  & \longrightarrow  & -i|g\rangle
_{m}|g\rangle _{n}|1\rangle  & \longrightarrow  & i|g\rangle _{m}|g\rangle
_{n}|1\rangle  \\
|e_{0}\rangle _{m}|e_{0}\rangle _{n}|0\rangle  & \longrightarrow  &
-i|g\rangle _{m}|e_{0}\rangle _{n}|1\rangle  & \longrightarrow  &
-i|g\rangle _{m}|r_{0}\rangle _{n}|1\rangle
\end{array}%
\end{equation*}%
\begin{equation}
\begin{array}[b]{rr}
\hat{U}_{m}^{1,0} &  \\
\longrightarrow  & |g\rangle _{m}|g\rangle _{n}|0\rangle , \\
\longrightarrow  & |g\rangle _{m}|e_{0}\rangle _{n}|0\rangle , \\
\longrightarrow  & |e_{0}\rangle _{m}|g\rangle _{n}|0\rangle , \\
\longrightarrow  & -|e_{0}\rangle _{m}|e_{0}\rangle _{n}|0\rangle .%
\end{array}
\label{bigone}
\end{equation}%
The effect of this interaction is to change the sign of the state only when
both ions are initially excited. Note that the state of the CM mode is
restored to the vacuum state $|0\rangle $ after the process. Equation (\ref%
{bigone}) is phase gate $|\epsilon _{1}\rangle |\epsilon _{2}\rangle
\rightarrow (-1)^{\epsilon _{1}\epsilon _{2}}|\epsilon _{1}\rangle |\epsilon
_{2}\rangle $ ($\epsilon _{1,2}=0,1$) which together with single qubit
rotations becomes equivalent to a controlled-NOT.

\subsection{Fast and robust 2-qubit gates for scalable ion trap quantum
computing}

Scalability of ion trap quantum computing is based on storing a set of ions
in a memory area, and moving ions independently to a processing unit: in
particular one must bring together \emph{pairs of ions} to perform a
two-qubit gate \cite{Wineland+MonroeETAL-NISTbible:98,
Cirac+Zoller-ScalableQCTrappedIons:00,
Kielpinksi+Monroe+Wineland-ScalableQC:02}. Basic steps towards this goal
have already been demonstrated experimentally \cite%
{Leibfried+DeMarcoETAL-NISTreview:03}. An important question to be addressed
is to identify the current limitations of the two--qubit gates with trapped
ions (given the fact that one--qubit gates are significantly simpler with
those systems). The ideal scheme should\cite%
{Steane+RoosETAL-Speeion-quanproc:00}:

\begin{description}
\item[(i)] be independent of temperature, so that one does not need to cool
the ions to their ground state after they are moved to or from their storage
area);

\item[(ii)] require no addressability (to allow the ions to be as close as
possible during the gate to increase their interaction strength), and

\item[(iii)] be fast, in order to minimize the effects of decoherence during
the gate, and to speed up the computation.
\end{description}

This last property has been identified as a key limitation\cite%
{Levi-PhysTodayIonTrapQC:03}: in essentially all schemes suggested so far
\cite{Cirac+Zoller-QuantCompColdTrappedIons:95,
Sorensen+Molmer-EntanglementQCIonsThermalMotion:00,
Sorensen+Molmer-PRLQCIonThermalMotion:99,
Knight00,Milburn+SchneiderETAL-trapquancompwith:00} one has to resolve
spectroscopically the motional sidebands of the ions with the exciting
laser, which limits the laser intensity and therefore the gate time. The
coherent control-gate gate between pairs of ions \cite%
{Garcia-Ripoll+Zoller+Cirac-SpeedOptimizedIonGate:03} analyzed below
overcomes this problem by not using spectral methods to couple the ion
motion to the internal states but rather mechanical effects.

As our model we consider two ions in a one--dimensional harmonic trap,
interacting with a laser beam on resonance. The Hamiltonian describing this
situation can be written as $H=H_{0}+H_{1}$, where $H_{0}=\nu_{c}a^{\dagger
}a+\nu_{r}b^{\dagger}b$ describes the motion in the trap and
\begin{equation}  \label{hamil}
H_{1}=\frac{1}{2}\Omega(t)\left[ \sigma_{1}^{+}e^{i\eta _{c}(a^{\dagger}+a)+%
\tfrac{1}{2}\eta_{r}(b^{\dagger}+b)}+\sigma_{2}^{+}e^{i\eta_{c}(a^{%
\dagger}+a)-\tfrac{1}{2}\eta_{r}(b^{\dagger}+b)}\right] +\text{h.c.}
\end{equation}
Here, $a$ and $b$ are the annihilation operators center-of-mass and
stretching mode, respectively, and $\nu_{c}=\nu$ and $\nu_{r}=\sqrt{3}%
\nu_{c} $ the corresponding frequencies. We denote by $\eta_{c}=\eta/\sqrt{2}
$ and $\eta_{r}=\eta\sqrt[4]{4/3}$ are to associated Lamb--Dicke parameters.
Note that the Rabi frequency $\Omega$ is the same for both ions, i.e. we
have not assumed individual addressing.

In the following we will consider two different kind of processes:

\begin{description}
\item[(i)] Free evolution, in which the laser is switched off ($\Omega=0$)
for a certain time;

\item[(ii)] Sequences of pairs of very fast laser pulses, each of them
coming from opposite sides, with duration $\delta t$ long enough to form a $%
\pi $-pulse ($\Omega\delta t = \pi$), but very short compared to the period
of the trap ($\nu\delta t \ll1$).
\end{description}

\begin{figure}[t]
\includegraphics[width=6cm]{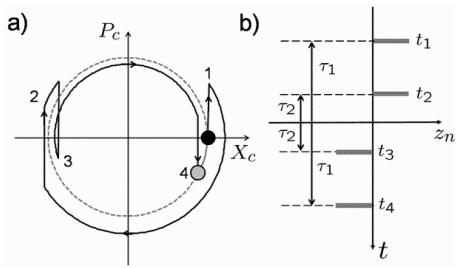}
\caption{a) Trajectory in phase space of the center-of-mass state of the ion
$(X_{c},P_{c})$ (where $(X_{c}+iP_{c})/\protect\sqrt{2}=\langle a \rangle$)
during the 2-qubit gate (solid line), connecting the initial state (black
filled circle) to the final state (grey filled circle) at the gate time $T$.
The time evolution consists of a sequence of kicks (vertical displacements),
which are interspersed with free harmonic oscillator evolution (motion along
the arcs). A pulse sequence satisfying the commensurability condition (%
\protect\ref{cond}) guarantees that the final phase space point is restored
to the one corresponding to a free harmonic evolution (dashed circle). The
particular pulse sequence plotted corresponds to a four pulse sequence given
in the text (Protocol I). Figure b) shows how the laser pulses (bars)
distribute in time for this scheme.}
\label{fig-1}
\end{figure}

Processes (i) and (ii) will be alternated: at time $t_{1}$ a sequence of $%
z_{1}$ pulses is applied, followed by free evolution until at time $t_{2}$
another sequence of $z_{2}$ pulses is applied followed by free evolution and
so on. The numbers $z_{k}$ are integers, whose sign indicates the direction
of the laser pulses. We can visualize the motion of the ions as a trajectory
in phase space. This is illustrated in Fig.~\ref{fig-1} for the
center-of-mass state of a single ion $(X_{c},P_{c})$, where $(X_{c}+iP_{c})/%
\sqrt{2}=\langle a\rangle$. The time evolution consists of a sequence of
kicks (vertical displacements), which are interspersed with free harmonic
oscillator evolution (motion along the arcs). The question is now whether we
can find a pulse sequence, such that the final phase space point (solid
line) is \emph{restored} to the one corresponding to a \emph{free} harmonic
evolution (dashed circle). In an appendix at the end of this section we show
that this can be achieved if the pulse sequence satisfies a \emph{%
commensurability condition} for the center-of-mass and stretch-mode
\begin{equation}  \label{cond}
C_{c}\equiv\sum_{k=1}^{N}z_{k}e^{-i\nu t_{k}}=0,\quad
C_{r}\equiv\sum_{k=1}^{N}z_{k}e^{-i\sqrt{3}\nu t_{k}}=0\;.
\end{equation}
In this case, the motional state of the ion will not depend on the qubits.
Thus the evolution operator is given by (see appendix)
\begin{equation}
\mathcal{U}(\Theta)=e^{i\Theta\sigma_{1}^{z}\sigma_{2}^{z}}e^{-i\nu
_{c}Ta^{\dagger}a}e^{-i\nu_{r}Tb^{\dagger}b},   \label{evolve}
\end{equation}
where $T$ is the total time required by the gate and
\begin{equation}
\Theta=4\eta^{2}\sum_{m=2}^{N}\sum_{k=1}^{m-1}z_{k}z_{m}\left[ \frac {\sin[%
\sqrt{3}\nu\Delta t_{km}]}{\sqrt{3}}-\sin(\nu\Delta t_{km})\right] ,
\label{phase}
\end{equation}
is a function of the spacing between laser pulses $\Delta t_{km}=t_{k}-t_{m}$%
. Therefore, if (\ref{cond}) are fulfilled, and $\Theta=\pi/4$ we will
produce a controlled--phase gate (which is equivalent to a controlled--NOT
gate up to local operations) which is \emph{completely independent of the
initial motional state}, i.e. there are no temperature requirements.

It can be shown \cite{Garcia-Ripoll+Zoller+Cirac-SpeedOptimizedIonGate:03}
that for any value of the time $T$ it is always possible to find a sequence
of laser pulses which implements the gate, and therefore the gate operation
can be, in principle, arbitrarily fast. We give two simple protocols.

\emph{Protocol I:} This protocol (see Fig. \ref{fig-1}) requires the least
number of pulses and produces the gate in a fixed time $T\simeq1.08(2\pi/\nu
)$. The sequence of pulses is defined as
\begin{equation}
(z_{n}/N, t_{n})= \{(\gamma,-\tau_{1}), (1,-\tau_{2}), (-1,\tau_{2}),
(-\gamma,\tau_{1})\}.
\end{equation}
Here $0 < \gamma= \cos(\theta) < 1.0$ is a real number, which may be
introduced by tilting both lasers a small angle $\theta$ with respect to the
axis of the trap, so that no transverse motion is excited. It is always
possible to find a solution to Eq.~(\ref{cond}) with $\tau_{1} \simeq
0.538(4)(2\pi/\nu) > \tau_{2} > 0$.

\begin{figure}[ptb]
\includegraphics[width=6cm]{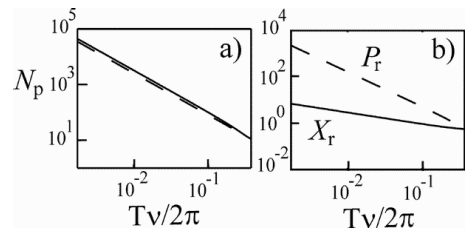}
\caption{(a) Log-log plot of the number of pairs of pulses required to
produce a phase gate using protocol II, as a function of the duration of the
gate, $T$, for a realistic value\protect\cite%
{DeMarco+Ben-KishETAL-Expedemocontwave:02} of the Lamb--Dicke parameter, $%
\protect\eta=0.178$. We plot both the exact result (solid line) and a rough
estimate $N_{P} = 40 (\protect\nu T/2\protect\pi)^{-3/2}$ (dashed line)
based on perturbative calculations. (b) Maximum relative displacement, $X_{r}
$ (solid), and maximum momentum acquired, $P_{r}$ (dashed line), for scheme
II. These quantities are dimensionless (scaled) versions of the real
observables, $X_{r}=\max[\langle x_{r}(t)\rangle/a_{0}]$, and $P_{r}\max[%
\langle p_{r}(t) \rangle a_{0}/\hbar]$.}
\label{fig-2}
\end{figure}

\emph{Protocol II:} This protocol performs the gate in an arbitrarily short
time $T$. The pulses are now distributed according to
\begin{eqnarray}
(z_{n}/N,t_{n}) &=&\{(-2,-\tau _{1}),(3,-\tau _{2}),(-2,-\tau _{3})  \notag
\\
&&,(2,\tau _{3}),(-3,\tau _{2}),(2,\tau _{1})\}.
\end{eqnarray}%
The whole process takes a time $T=2\tau _{1}$ and requires $N_{p}=\sum
|z_{n}|=14N$ pairs of pulses. As Fig.~\ref{fig-2} shows, the number of
pulses increases with decreasing time as $N_{p}\propto T^{-3/2}$.

Ref.~\cite{Garcia-Ripoll+Zoller+Cirac-SpeedOptimizedIonGate:03} gives a
detailed study of the main limitations of the the scheme, and provides
quantitative estimates for the gate fidelity. On the list of imperfections
is first of all anharmonicities of the restoring forces. The more pulses we
apply, the larger the relative displacement of the ions, as Fig.~\ref{fig-2}%
(b) shows. When the ions become too close to each other, the increasing
intensity of the Coulomb force can lead to a breakdown of the harmonic
approximation which is implicit in Eq.~(\ref{hamil}). Imposing an error $%
E\simeq10^{-4}$ we estimate the shortest realistic time to be $\nu
T\simeq10^{-3}$ \cite{Garcia-Ripoll+Zoller+Cirac-SpeedOptimizedIonGate:03}.
In addition, laser pulses have a finite duration. However, even for
relatively long pulses, we obtain a fidelity which is comparable to the
results obtained in current setups \cite%
{Schmidt-Kaler+BlattETAL-CiracZollerGate:03,Leibfried+DeMarcoETAL-ExpDemTwoIonPhaseGate:03}%
. As mentioned before, the scheme is also insensitive to temperature. If the
commensurability condition (\ref{cond}) is not perfectly satisfied due to,
for example, errors in timing of laser pulses, or misalignment of the
lasers, then the corresponding contribution to the gate error is still a
weak function of temperature.

Finally, we remark that it is not necessary to kick the atoms using pairs of
counter-propagating laser beams. The same effect (i.e. a change of sign in $%
\eta$) may also be achieved in current experiments by reverting the internal
state of both ions simultaneously. One then only needs a laser beam (aligned
with the trap) to kick the atoms, and another laser (orthogonal to the axis
of the trap) to produce the NOT gate. The second and more important remark
is that it is possible to avoid errors in the laser pulses by using an
adiabatic passage scheme (see references cited in \cite%
{Cirac+Duan+Jaksch+Zoller-Varenna:02}) which is insensitive to fluctuations
in the laser intensity. In addition, this method also tolerates that the two
ions see slightly different laser intensity.

In summary, the new concept of a ``coherent control'' two-qubit quantum gate
allows operations on a time scale up three orders of magnitude faster than
the trap frequency, while at the same time requiring no single ion
addressing, no Lamb-Dicke assumption, and ground state cooling of the ion,
and being robust against imperfections.

\emph{Appendix: Derivation of Eqs.~(\ref{cond}) and (\ref{evolve}). } Here
we presents details of the derivation of the commensurability condition (\ref%
{cond}) to achieve the factorization of the motional states according to
Eq.~(\ref{evolve}). For a pulse sequence, consisting of kicks interspersed
with free harmonic time evolution (Fig.~\ref{fig-1}), we write $\mathcal{U}=%
\mathcal{U}_{c} \mathcal{U}_{r}$, where $\mathcal{U}_{c,r}= \prod_{k=1}^{N}
U_{c,r} (\Delta t_{k},z_{k})$ has contributions for center-of-mass and
relative motion, 
\begin{align*}
U_{c}(t_{k},z_{k}) & = e^{-i 2z_{k}\eta_{c}(a+a^{\dagger})(\sigma_{1}^{z}+
\sigma_{2}^{z})} e^{-i\nu_{c} \Delta t_{k} a^{\dagger}a}, \\
U_{r}(t_{k},z_{k}) & = e^{-i z_{k}\eta_{r}(b+b^{\dagger})(\sigma_{1}^{z}-
\sigma_{2}^{z})} e^{-i\nu_{r} \Delta t_{k} b^{\dagger}b}.
\end{align*}
The integers $z_{k}$ indicate the direction of the initial pulse in the
sequence of pairs of very fast laser pulses, each of them coming from
opposite sites.

In order to fully characterize $U$, we only have to investigate its action
on states of the form $|i\rangle_{1}|j\rangle_{2} |\alpha\rangle_{c}|\beta
\rangle_{r} $, where $i,j=0,1$ denote the computational basis, and $%
|\alpha\rangle$ and $|\beta\rangle$ are coherent states. This task can be
easily carried out once we know the action of $\mathcal{U}=\prod_{k=1}^{N}
U(\phi_{k},p_{k})$ on an arbitrary coherent state $|\alpha\rangle$, where
\begin{equation*}
U(\phi_{k},p_{k}) = e^{-i p(a+a^{\dagger})} e^{-i\phi_{k} a^{\dagger}a}.
\end{equation*}
We obtain $\mathcal{U}|\alpha\rangle= e^{i\xi} |\tilde\alpha\rangle$, where
\begin{align*}
\tilde\alpha & = \alpha e^{-i\theta_{N}} - i \sum_{k=1}^{N} p_{k}
e^{i(\theta_{k}-\theta_{N})}, \\
\xi & = - \sum_{m=2}^{N}\sum_{k=1}^{m-1} p_{m}p_{k}
\sin(\theta_{k}-\theta_{m}) - \Re\left[ \alpha\sum_{k=0}^{N} p_{k}
e^{-i\theta_{m}} \right] ,
\end{align*}
with $\theta_{k}=\sum_{m=1}^{k} \phi_{m}$.

The crucial point is to realize that if $\sum_{k=1}^{N} p_{k}
e^{i\theta_{k}}=0$ the motional state $|\alpha\rangle$ after the evolution
is the same as if there was only free evolution (Fig.~1a), and a global
phase $\xi$ appears which does not depend on the motional state (Fig.~1a).
Translating this result to the operators $\mathcal{U}_{c}|\alpha\rangle$ and
$\mathcal{U}_{r}|\beta\rangle$, we obtain condition (\ref{cond}) for Eq.~(%
\ref{evolve}) to be valid.

\section{Atoms in optical lattices}

\label{sec:coldatoms}

Bose Einstein condensates (BEC) are a source of a large number of ultracold
atoms and, as we will show below, they can also be developed as a tool to
provide a \emph{large} number of qubits stored in optical lattices. In a
condensate, due to the weak interactions, all atoms occupy the single
particle ground state of the trapping potential, corresponding to a product
state of the wave function.

This picture is must be revised by inducing a degeneracy in the ground state
which is comparable to the number of atoms. For instance, as first proposed
in Refs.~\cite%
{Jaksch+BruderETAL-ColdBosonicAtomOpticalLattices:98,JakschETAL-EntanglementColdCollision:99}%
, it is possible to load a BEC in a deep 3D optical lattice forming a
perfect Mott insulator phase with one atom per lattice site. The system is
no longer a BEC, but an array of a large number of identifiable qubits, that
can be entangled in massively parallel operation with spin-dependent
lattices \cite{JakschETAL-EntanglementColdCollision:99}. This scenario has
recently been realized in the laboratory in a series of remarkable
experiments in Munich \cite{Bloch-SF-Mott:02,Bloch-CCEntanglement:03}.
Entanglement of atoms in a lattice can also be achieved by dipole-dipole
interactions \cite%
{Brennen+Caves-Deutsch-QuantumLogicGateOpticalLattice:99,Brennen+Deutsch-Williams-QuanLogiTrapAtom:02}%
), and the interactions and the speed of the quantum operations may be
significantly enhanced using the very strong interactions are obtained
between laser excited Rydberg states \cite{JakschETAL-RydbergGate:00}.

\subsection{Cold atoms in optical lattices: the Hubbard model}

Optical lattices are periodic arrays of microtraps for cold atoms generated
by standing wave laser fields. The periodic structure of the lattice gives
rise to a series of Bloch bands for the atomic center-of-mass motion. Atoms
loaded in an optical lattice from a BEC will only occupy the lowest Bloch
band due to the low temperatures. The physics of these atoms can be
understood in terms of a Hubbard model with Hamiltonian \cite%
{Jaksch+BruderETAL-ColdBosonicAtomOpticalLattices:98}
\begin{equation}  \label{Hubbard}
H=-\sum_{\langle i,j\rangle}J_{ij}b_{i}^{\dag}b_{j}+\frac{1}{2}%
U\sum_{i}b_{i}^{\dag}b_{i}^{\dag}b_{i}b_{i} \; .
\end{equation}
Here $b_{i}$ and $b_{i}^{\dag}$ are bosonic destruction operators for atoms
at each lattice site satisfying the bosonic commutation relations $[ b_{i},
b_{j}^{\dagger}] = \delta_{ij}$. The tunneling of the atoms between
different sites is described by the hopping matrix elements $J_{ij}$. The
parameter $U$ is the onsite interaction of atoms resulting from the
collisional interactions. The distinguishing feature of this system is the
time dependent control of the parameters $J_{ij}$ (kinetic energy) and $U$
(potential energy) by the intensity of the lattice laser. Increasing the
intensity of the laser deepens the lattice potential, and suppresses the
hopping while at the same time increasing the atomic density at each lattice
site and thus the onsite interaction. For shallow lattices $J_{ij} \gg U$
the kinetic energy is dominant, and the ground state of $N$ atoms will be a
superfluid in which all bosonic atoms occupy the lowest momentum state in
the Bloch band, $(\sum _{i}b^{\dagger})^{N} |\text{vac}\rangle$. If $J_{ij}
\ll U$, on the other hand, the interactions dominate: for commensurate
filling, i.e.~when the number of lattice site matches the number of atoms,
the ground state becomes a Mott-insulator state $b_{1}^{\dagger}\ldots b_{N}
^{\dagger}|\text{vac}\rangle$ (Fock state of atoms). The
superfluid-Mott-insulator transition is an example of a so-called quantum
phase transition as studied in Ref.~\cite{Sachdev-QuantPhaseTransition:99}.
This Mott insulator regime is of particular interest, as it provides a \emph{%
very large} number of identifiable atoms located in the the array of
microtraps provided by the optical lattice, whose internal hyperfine or spin
states can serve as qubits\cite%
{Jaksch+BruderETAL-ColdBosonicAtomOpticalLattices:98}. The first
experimental realization of the Mott insulator quantum phase transition was
recently reported by Bloch and collaborators \cite{Bloch-SF-Mott:02}.

\begin{figure}[ptb]
\includegraphics[width=6.5cm]{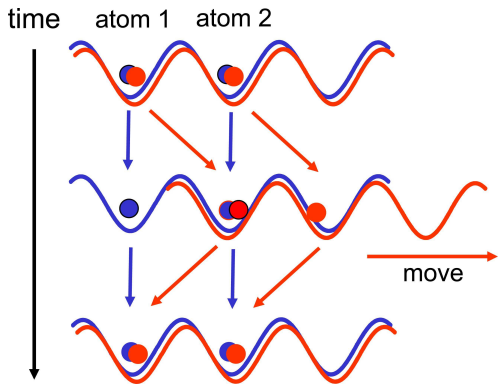}
\caption{ Controlled collisions of two atoms with internal states $|0\rangle$
and $|1\rangle$ (red and blue circles) in a moveable state-dependent optical
lattice (red and blue lattice) to entangle two atoms\protect\cite%
{JakschETAL-EntanglementColdCollision:99,Bloch-CCEntanglement:03}. This
scheme unterlies the quantum simulator on the optical lattice.}
\label{lattice1}
\end{figure}

\begin{figure}[ptb]
\includegraphics[width=3.5cm]{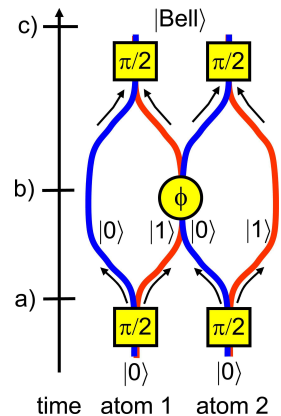}
\caption{Ramsey experiment with two atoms colliding in a lattice to generate
a Bell state following Ref.~\protect\cite%
{JakschETAL-EntanglementColdCollision:99,Bloch-CCEntanglement:03}. Time
evolution is from bottom to top. The two atoms are initially prepared in the
product state $|0\rangle|0\rangle$. A $\protect\pi/2$ pulse generates the
(unnormalized) superposition state $(|0\rangle+|1\rangle)((|0\rangle
+|1\rangle))$ (a). A coherent collision provides a phase shift $\protect\phi$
conditional to the first atom being in state $|1\rangle$ and the second atom
being in state $|0\rangle$, i.e. $|0\rangle|0\rangle+ e^{i \protect\phi%
}|0\rangle|1\rangle+|1\rangle|0\rangle+|1\rangle|1\rangle$ (b). A final $%
\protect\pi /2$-pulse closes the Ramsey interferometer resulting in the
state $(1-e^{i\protect\phi})|\text{Bell}\rangle+(1+e^{i\protect\phi%
}|1\rangle|1\rangle$, which for $\protect\phi=\protect\pi$ is a Bell state.}
\label{lattice2}
\end{figure}

\subsection{Entanglement via coherent ground state collisions}

Entanglement of qubits represented by cold atoms in a Mott-phase can be
obtained by combing the collisional interactions (compare the onsite
interaction in Eq.~\ref{Hubbard}) with a \emph{spin-dependent} optical
lattice \cite{JakschETAL-EntanglementColdCollision:99}. Let us assume that
qubits $|0\rangle$, $|1\rangle$ are stored in two longlived atomic hyperfine
ground states. With an appropriate choice of atomic states and the laser
configurations \cite{JakschETAL-EntanglementColdCollision:99} we can
generate an optical lattice which is spin-dependent, i.e. atoms in $|0\rangle
$ and $|1\rangle$ see a different optical potential. In addition these two
optical potentials can change in time, so that both lattices have a tunable
separation.

This provides us with a mechanism to \emph{move} atoms \emph{conditional} to
the state of the qubit. In particular, we can \emph{collide} two atoms
\textquotedblleft by hand\textquotedblright\ , as illustrated in Fig.~\ref%
{lattice1}, so that only the component of the wave function with the first
atom in $|1\rangle$ and the second atom in $|0\rangle$ will pick up a
collisional phase $\phi$, which entangles the atoms. In fact, this
interaction gives rise to a phase gate between adjacent atoms $%
|1\rangle_{i}|0\rangle_{i+1}\longrightarrow e^{i\phi}
|1\rangle_{i}|0\rangle_{i+1}$. In Fig.~\ref{lattice2} we illustrate a Ramsey
type experiment to generate and detect a Bell state via these collisional
interactions. Again this idea has been demonstrated recently in a seminal
experiment in the Munich group\cite{Bloch-CCEntanglement:03}. This
conditional quantum logic can also be realized with magnetic, electric
microtraps and microoptical dipole traps \cite%
{CalarcoETAL-QuantumGateCollisions:00,Schlosser+Grangier-TwoAtomTrap:01,Buchkremer+Dumke+Birkl+Ertmer-QIMicrofabricatedOpticalElements:02}%
.

In more detail, we consider a situation where two atoms with electrons
populating the internal states $|0\rangle$ and $|1\rangle$, respectively,
are trapped in the ground states $\psi_{0}^{0,1}$ of two potential wells $%
V^{0,1}$. Initially, these wells are centered at positions $\bar{x}^{0}$ and
$\bar
{x}^{1}$, sufficiently far apart (distance $d=\bar{x}_{1}-\bar{x} _{0}
$) so that the particles do not interact. The positions of the potentials
are moved along trajectories $\bar{x}^{0}(t)$ and $\bar{x}^{1}(t)$ so that
the wavepackets of the atoms overlap for certain time, until finally they
are restored to the initial position at the final time. This situation is
described by the Hamiltonian
\begin{equation}
H\!=\!\sum_{\beta=0,1}\left[ \frac{(\hat{p}^{\beta})^{2}}{2m}+V^{\beta
}\left( \hat{x}^{\beta}\!-\!\bar{x}^{\beta}(t)\right) \right] +u^{\mathrm{01}%
}(\hat{x}^{0}\!-\!\hat{x}^{1}).   \label{Hamil}
\end{equation}
Here, $\hat{x}^{0,1}$ and $\hat{p}^{0,1}$ are position and momentum
operators, $V^{0,1}\left( \hat{x}^{0,1}-\bar{x}^{0,1}(t)\right) $ describe
the displaced trap potentials and $u^{\mathrm{01}}$ is the atom--atom
interaction term. Ideally, we would like to implement the transformation
from before to after the collision,
\begin{equation}
\psi_{0}^{0}(x^{0}\!-\!\bar{x}^{0})\psi_{0}^{1}(x^{b}\!-\!\bar{x}%
^{1})\rightarrow e^{i\phi}\psi_{0}^{0}(x^{0}\!-\!\bar{x}^{0})%
\psi_{0}^{1}(x^{1}\!-\!\bar{x}^{1}),   \label{transf}
\end{equation}
where each atom remains in the ground state of its trapping potential and
preserves its internal state. The phase $\phi$ will contain a contribution
from the interaction (collision). The transformation (\ref{transf}) can be
realized in the \emph{adiabatic limit}, whereby we move the potentials
slowly on the scale given by the trap frequency, so that the atoms remain in
the ground state. Moving non-interacting atoms will induce kinetic single
particle kinetic phases. In the presence of interactions ($u^{\mathrm{ab}%
}\neq0$), we define the time--dependent energy shift due to the interaction
as
\begin{equation}
\Delta E(t)=\frac{4\pi a_{s}\hbar^{2}}{m}\int dx|\psi_{0}^{0}\left( x-\bar {x%
}^{0}(t)\right) |^{2}|\psi_{0}^{1}\left( x-\bar{x}^{1}(t)\right) |^{2},
\label{deltaE}
\end{equation}
where $a_{s}$ is the $s$--wave scattering length. We assume that $|\Delta
E(t)|\ll\hbar\nu$ with $\nu$ the trap frequency so that no sloshing motion
is excited. In this case, (\ref{transf}) still holds with $%
\phi=\phi^{0}+\phi ^{1}+\phi^{\mathrm{01}}$, where in addition to (trivial)
single particle kinetic phases $\phi^{0}$ and $\phi^{1}$ arising from moving
the potentials, we have a \emph{collisional phase shift}
\begin{equation}
\phi^{\mathrm{01}}=\int_{-\infty}^{\infty}dt\Delta E(t)/\hbar.
\label{phicol}
\end{equation}

If the first atom is in a superposition state of the two qubits, the atomic
wave packet would be ''split'' by moving the state dependent potentials,
very much like with a beam splitter in atom interferometry. Thus we can move
the potentials of neighboring atoms such that only the $|0\rangle$ component
of the first atom ``collides'' with the state $|0\rangle$ of the second atom
\begin{align}
|0\rangle_{1}|0\rangle_{2} & \rightarrow e^{i2\phi^{0}}|0\rangle
_{1}|0\rangle_{2},  \notag  \label{gate} \\
|0\rangle_{1}|1\rangle_{2} & \rightarrow e^{i(\phi^{0}+\phi^{1}+\phi^{%
\mathrm{01}})}|0\rangle_{1}|1\rangle_{2},  \notag \\
|1\rangle_{1}|0\rangle_{2} & \rightarrow
e^{i(\phi^{0}+\phi^{1})}|b\rangle_{1}|0\rangle_{2},  \notag \\
|1\rangle_{1}|1\rangle_{2} & \rightarrow e^{i2\phi^{1}}|1\rangle
_{1}|1\rangle_{2},
\end{align}
where the motional states remain unchanged in the adiabatic limit, and $%
\phi^{0}$ and $\phi^{1}$ are single particle kinetic phases. The
transformation (\ref{gate}) corresponds to a fundamental two--qubit gate.
The fidelity of this gate is limited by nonadiabatic effects, decoherence
due to spontaneous emission in the optical potentials and collisional loss
to other unwanted states, or collisional to unwanted states. According to
Ref.~\cite{JakschETAL-EntanglementColdCollision:99} the fidelity of this
gate operation is remarkably close to one in a large parameter range.

\subsection{Application: quantum simulations}

Applying the previous method to an optical lattice that has more qubits, we
can entangle many atoms with a single lattice movement, i.e.~in a highly
parallel entanglement operation. While for two atoms we have obtained a Bell
state (see Fig.~\ref{lattice2}), for three atoms this produces a maximally
entangled GHZ-state, and for $2D$ lattices this allows the generation of a
cluster state, which is the basic resource for universal quantum computing
in Briegel \emph{et al.}'s one way quantum computer\cite%
{Raussendorf+Briegel-OneWayQuantumComputer:01}.

The parallelism inherent in the lattice movements makes \textquotedblleft
atoms in optical lattices\textquotedblright\ an ideal candidate for a
Feynman-type quantum simulator (see the Appendix of this section) for
bosonic, fermionic and spin many body systems, allowing simulation of
various types and strengths of particle interactions, and $1,2$ or $3D$
lattice configurations in a regime of many atoms, clearly unaccessible to
any classical computer. By a stroboscopic switching of laser pulses and
lattice movements combined with collisional interactions one can implement
sequences of 1 and 2-qubit operations to simulate the time evolution
operator of a many body system \cite{Jane+VidalETAL-Simulation:03}. For
translationally invariant systems, there is no need to address individual
lattice sites, which makes the requirements quite realistic in the light of
the present experimental developments. On the other hand, as noted above,
Hubbard Hamiltonians with interactions controlled by lasers can also be
realized directly with cold bosonic or fermionic atoms in optical lattices.
This \textquotedblleft analogue\textquotedblright\ quantum simulation
provides a direct way of studying properties of strongly correlated systems
in cold atom labs, which in the future may develop into a novel tool of
condensed matter physics.

For the near future, we expect that atoms in optical lattices will be used
to simulate a variety of other physical systems like, for example,
interacting Fermions in 2 Dimensions using different lattice geometries. We
also expect an important progress towards loading single (neutral) atoms in
different types of potentials (optical, magnetic, etc), and the performance
of quantum gates with few of these systems. This would allow to create few
atom entangled states which may be used to observe violations of Bell
inequalities, or to observe interesting phenomena like teleportation or
error correction. As opposed to the trapped ions, at the moment it is hard
to predict whether scalable quantum computation will be possible with
neutral atoms in optical lattices using the present experimental set--ups.
In any case, due to the high parallelism of these systems, we can clearly
foresee that they will allow us to obtain a very deep insight in condensed
matter physics via quantum simulations.

\emph{Appendix: Quantum simulator} In brief, the basic concept of the
quantum simulator is as follows. Let us consider a quantum system composed
of $N$ qubits all initially in state $|0\rangle$. We apply a two--qubit gate
(specified by a $4\times4$ unitary matrix) to the first and second qubit,
another one to the second and the third, and so on until we have performed $%
N-1$ such gates. Now, we measure the last qubit in the basis $|0\rangle
,|1\rangle$. Let us denote by $p_{0}$ and $p_{1}$ the probability of
obtaining $0$ and $1$ in this measurement. Our goal is to determine such
probabilities with a prescribed precision (for example, of 1\%). A way to
determine the probabilities using a classical computer is to simulate the
whole process: we take a vector which has $2^{N}$ components and multiply it
by a $2^{N}\times2^{N}$ matrix every time we simulate the action of a gate.
At the end we can calculate the desired probabilities using the standard
rules of Quantum Mechanics. However, as soon as $N$ is of the order of 30,
we will not be able to store the vector and the matrices in any existing
computer. Moreover, the time required to simulate the action of the gates
will increase exponentially with the number of qubits. However, with a
quantum computer this simulation will required to repeat the same
computation of the order of $100$ times, and each computation requires only $%
N-1$ gates. Thus, we see that the quantum computer itself is much more
efficient to simulate quantum systems, something that Feynman already
pointed out in 1982 \cite%
{Nielsen+Chuang-QC:00,Lloyd-UniversalQuantumSimulators:96}. Of course, this
particular example is artificial, and it is not related to a real problem.
However, there exist physical systems which cannot be simulated with
classical computers but in which a quantum computer could offer an important
insight on some physical phenomena which are not yet understood \cite%
{Lloyd-UniversalQuantumSimulators:96}. For example, one could use a quantum
computer to simulate spin systems or Hubbard models, and extract some
information about open questions in condensed matter physics. Another
possibility is to use an "analogue" quantum computer (as our artificial
Hubbard models) to do the job, i.e. to choose a system which is described by
the same Hamiltonian which one wants to simulate, but that can be very well
controlled and measured.

\section{Quantum information processing with atomic ensembles}

\label{sec:atomicensembles}

\subsection{Introduction}

In the previous section, the quantum computation schemes are based on laser
manipulation of single trapped particles. Here, we will show that laser
manipulation of macroscopic atomic ensembles can also be exploited for
implementation of quantum information processing \cite%
{Lukin002,Duan01b,Kuzmich00,Duan002,FL00,Julsgaard01,Duan02,Kuzmich03,Lukin03,ustc03}%
. In particular, we will discuss the uses of this system for continuous
variable quantum teleportation and for implementation of quantum repeaters
which enable scalable long-distance quantum communication.

The atomic ensemble contains a large number of identical neutral atoms,
whose experimental candidates can be either laser-cooled atoms \cite%
{Hald99,Hau99,Kuzmich03}, or room-temperature gas \cite%
{Scully99,Kuzmich001,Julsgaard01,Lukin03,ustc03}. The motivation of using
atomic ensembles instead of single-particles for quantum information
processing is mainly two-folds: firstly, laser manipulation of atomic
ensembles without separate addressing of individual atoms is typically much
easier than the laser manipulation of single particles; secondly and more
importantly, the use of the atomic ensembles allows for some collective
effects resulting from many-atom coherence to enhance the signal-to-noise
ratio, which is critical for implementations of some quantum information
protocols.

In the next section, we first show the ideas of using atomic ensembles for
implementation of scalable long-distance quantum communication.
Long-distance quantum communication is necessarily based on the use of
photonic channels. However, due to losses and decoherence in the channel,
the communication fidelity decreases exponentially with the channel length.
To overcome this outstanding problem, one needs to use the concept of
quantum repeaters \cite{Briegel98}, which provide the only known way for
robust long-distance quantum communication. The best known method for
complete implementation of quantum repeaters with sensible experimental
technologies was proposed in Ref. \cite{Duan01b}. Significant experimental
advances haven been achieved recently towards realization of this
comprehensive scheme, and we will briefly review these advances. In the
final section, we discuss the use of atomic ensembles for continuous
variable quantum information processing. Laser manipulation of atomic
ensembles provides an elegant way for realizing continuous variable atomic
quantum teleportation \cite{Duan002}, and we will review the basic
theoretical schemes as well as the following experimental achievements.

\subsection{Atomic ensembles for implementation of quantum repeaters}

Quantum communication is an essential element required for constructing
quantum networks and for secretly transferring messages by means of quantum
cryptography. The central problem of quantum communication is to generate
nearly perfect entangled states between distant sites. Such states can be
used then to implement secure quantum cryptography \cite{Ekert91} or to
transfer arbitrary quantum messages \cite{Bennett93}. The schemes for
quantum communication need to be based on the use of the photonic channels.
To overcome the inevitable signal attenuation in the channel, the concept of
entanglement purification was invented \cite{Bennett96}. However,
entanglement purification does not fully solve the problem for long-distance
quantum communication. Due to the exponential decay of the entanglement in
the channel, one needs an exponentially large number of partially entangled
states to obtain one highly entangled state, which means that for a
sufficiently long distance the task becomes nearly impossible.

The idea of quantum repeaters was proposed to solve the difficulty
associated with the exponential fidelity decay \cite{Briegel98}. In
principle, it allows to make the overall communication fidelity very close
to the unity, with the communication time growing only polynomially with the
transmission distance. In analogy to fault-tolerant quantum computing \cite%
{Preskill}, the quantum repeater proposal is a concatenated entanglement
purification protocol for communication systems. The basic idea is to divide
the transmission channel into many segments, with the length of each segment
comparable to the channel attenuation length. First, one generates
entanglement and purifies it for each segment; the purified entanglement is
then extended to a longer length by connecting two adjacent segments through
entanglement swapping \cite{Bennett93}. After entanglement swapping, the
overall entanglement is decreased, and one has to purify it again. One can
continue the rounds of the entanglement swapping and purification until a
nearly perfect entangled states are created between two distant sites.

To implement the quantum repeater protocol, one needs to generate
entanglement between distant quantum bits (qubits), store them for
sufficiently long time and perform local collective operations on several of
these qubits. The requirement of quantum memory is essential since all
purification protocols are probabilistic. When entanglement purification is
performed for each segment of the channel, quantum memory can be used to
keep the segment state if the purification succeeds and to repeat the
purification for the segments only where the previous attempt fails. This is
essentially important for polynomial scaling properties of the communication
efficiency since with no available memory we have to require that the
purifications for all the segments succeeds at the same time; the
probability of such event decreases exponentially with the channel length.
The requirement of quantum memory implies that we need to store the local
qubits in the atomic internal states instead of the photonic states since it
is difficult to store photons for a reasonably long time. With atoms as the
local information carriers it seems to be very hard to implement quantum
repeaters since normally one needs to achieve the strong coupling between
atoms and photons with high-finesse cavities for atomic entanglement
generation, purification, and swapping \cite{Cirac97,Enk98}, which, in spite
of the recent significant experimental advances \cite%
{Ye99,Rempe02,Chapman03,Kimble03}, remains a very challenging technology.

To overcome this difficulty, a scheme was proposed in Ref. \cite{Duan01b} to
realize quantum repeaters based on the use of atomic ensembles. The laser
manipulation of the atomic ensembles, together with simple linear optics
devices and routine single-photon detection, do the whole work for
long-distance quantum communication. This scheme combines entanglement
generation, connection, and application, with built-in entanglement
purification, and as a result, it is inherently resilient to influence of
noise and imperfections. Here, we will first explain the basic ideas of this
theoretical proposal and then review the recent experimental advances.

\subsubsection{Entanglement generation}

To realize long-distance quantum communication, first we need to entangle
two atomic ensembles within the channel attenuation length. This
entanglement generation scheme is based on single-photon interference at
photodetectors, which critically uses the fault-tolerance property of the
photon detection \cite{Cab99} and the collective enhancement of the
signal-to-noise ratio available in a many-atomic ensemble under an
appropriate interaction configuration \cite{Duan02A}.

The system is a sample of atoms prepared in the ground state $\left|
1\right\rangle $ with the level configuration shown in Fig.~ \ref%
{d63reviewp4}. This sample is illuminated by a short, off-resonant laser
pulse that induces Raman transitions into the state $\left| 2\right\rangle $
( a hyperfine level in the ground-state manifold with a long coherence
time). We are particularly interested in the forward-scattered Stokes light
that is co-propagating with the laser. Such scattering events are uniquely
correlated with the excitation of the symmetric collective atomic mode $S$
given by $S\equiv\left( 1/\sqrt{N_{a}}\right) \sum_{i}\left| g\right\rangle
_{i}\left\langle s\right| $ \cite{Duan02A}, where the summation is taken
over all the atoms. In particular, an emission of the single Stokes photon
in a forward direction results in the state of atomic ensemble given by $%
S^{\dagger}|0_{a}\rangle$, where the ensemble ground state $\left|
0_{a}\right\rangle \equiv \bigotimes_{i}\left| 1\right\rangle _{i}$.

We assume that the light-atom interaction time is short so that the mean
photon number in the forward-scattered Stokes pulse is much smaller than $1$%
. One can assign an effective single-mode bosonic operator $a$ for this
Stokes pulse with the corresponding vacuum state denoted by $\left|
0_{p}\right\rangle $. The whole state of the atomic collective mode and the
forward-scattered Stokes mode can now be written in the following form \cite%
{Duan02A}
\begin{equation}
\left| \phi\right\rangle =\left| 0_{a}\right\rangle \left|
0_{p}\right\rangle +\sqrt{p_{c}}S^{\dagger}a^{\dagger}\left|
0_{a}\right\rangle \left| 0_{p}\right\rangle +o\left( p_{c}\right) ,
\label{19}
\end{equation}
where $p_{c}$ is the small excitation probability.

\begin{figure}[ptb]
\label{d63reviewp4}
\par
\begin{center}
\includegraphics[width=6.0cm]{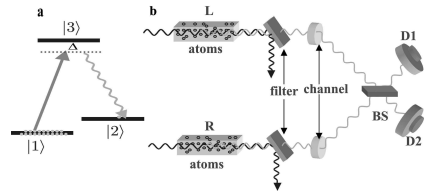}
\end{center}
\caption{(a) The relevant level structure of the atoms in the ensemble with $%
\left| 1\right\rangle $, the ground state, $\left| 2\right\rangle ,$ the
metastable state for storing a qubit, and $\left| 3\right\rangle ,$ the
excited state. The transition $\left| 1\right\rangle \rightarrow\left|
3\right\rangle $ is coupled by the classical laser with the Rabi frequency $%
\Omega$, and the forward scattering Stokes light comes from the transition $%
\left| 3\right\rangle \rightarrow\left| 2\right\rangle $. For convenience,
we assume off-resonant coupling with a large detuning $\Delta$. (b)
Schematic setup for generating entanglement between the two atomic ensembles
L and R. The two ensembles are pencil shaped and illuminated by the
synchronized classical laser pulses. The forward-scattering Stokes pulses
are collected after the filters (polarization and frequency selective) and
interfered at a 50\%-50\% beam splitter BS after the transmission channels,
with the outputs detected respectively by two single-photon detectors D1 and
D2. If there is a click in D1 \textit{or} D2, the process is finished and we
successfully generate entanglement between the ensembles L and R. Otherwise,
we first apply a repumping pulse to the transition $\left| 2\right\rangle
\rightarrow\left| 3\right\rangle $ on the ensembles L and R to set the state
of the ensembles back to the ground state $\left| 0\right\rangle
_{a}^{L}\otimes\left| 0\right\rangle _{a}^{R}$, then the same classical
laser pulses as the first round are applied to the transition $\left|
1\right\rangle \rightarrow\left| 3\right\rangle $ and we detect again the
forward-scattering Stokes pulses after the beam splitter. This process is
repeated until finally we have a click in the D1 \textit{or} D2 detector. }
\end{figure}

Now we explain how to use this setup to generate entanglement between two
distant ensembles L and R using the configuration shown in Fig.~ \ref%
{d63reviewp4}. Here, two laser pulses excited both ensembles simultaneously,
and the whole system is described by the state $\left| \phi\right\rangle
_{L}\otimes\left| \phi\right\rangle _{R}$, where $\left| \phi\right\rangle
_{L}$ and $\left| \phi\right\rangle _{R}$ are given by Eq. (\ref{19}) with
all the operators and states distinguished by the subscript L or R. The
forward scattered Stokes signal from both ensembles is combined at the beam
splitter and a photodetector click in either D1 \textit{or} D2 measures the
combined radiation from two samples, $a_{+}^{\dagger}a_{+}$ or $%
a_{-}^{\dagger}a_{-}$ with $a_{\pm}=\left( a_{L}\pm e^{i\varphi}a_{R}\right)
/\sqrt{2}$. Here, $\varphi$ denotes an unknown difference of the phase
shifts in the two-side channels. We can also assume that $\varphi$ has an
imaginary part to account for the possible asymmetry of the setup, which
will also be corrected automatically in our scheme. But the setup asymmetry
can be easily made very small, and for simplicity of expressions we assume
that $\varphi$ is real in the following. Conditional on the detector click,
we should apply $a_{+}$ or $a_{-}$ to the whole state $\left|
\phi\right\rangle _{L}\otimes\left| \phi\right\rangle _{R}$, and the
projected state of the ensembles L and R is nearly maximally entangled with
the form (neglecting the high-order terms $o\left( p_{c}\right) $)
\begin{equation}
\left| \Psi_{\varphi}\right\rangle _{LR}^{\pm}=\left( S_{L}^{\dagger}\pm
e^{i\varphi}S_{R}^{\dagger}\right) /\sqrt{2}\left| 0_{a}\right\rangle
_{L}\left| 0_{a}\right\rangle _{R}.   \label{20}
\end{equation}
The probability for getting a click is given by $p_{c}$ for each round, so
we need repeat the process about $1/p_{c}$ times for a successful
entanglement preparation, and the average preparation time is given by $%
T_{0}\sim t_{\Delta}/p_{c}$. The states $\left| \Psi_{r}\right\rangle
_{LR}^{+}$ and $\left| \Psi_{r}\right\rangle _{LR}^{-}$ can be easily
transformed to each other by a simple local phase shift. Without loss of
generality, we assume in the following that we generate the entangled state $%
\left| \Psi _{r}\right\rangle _{LR}^{+}$.

The presence of noise will modify the projected state of the ensembles to
\begin{equation}
\rho_{LR}\left( c_{0},\varphi\right) =\frac{1}{c_{0}+1}\left( c_{0}\left|
0_{a}0_{a}\right\rangle _{LR}\left\langle 0_{a}0_{a}\right| +\left|
\Psi_{\varphi}\right\rangle _{LR}^{\text{ }+}\left\langle \Psi_{\varphi
}\right| \right) ,   \label{21}
\end{equation}
where the ``vacuum'' coefficient $c_{0}$ is determined by the dark count
rates of the photon detectors. It will be seen below that any state in the
form of Eq. (\ref{21}) will be purified automatically to a maximally
entangled state in the entanglement-based communication schemes. We
therefore call this state an effective maximally entangled (EME) state with
the vacuum coefficient $c_{0}$ determining the purification efficiency.

\subsubsection{Entanglement connection through swapping}

\begin{figure}[ptb]
\label{d63reviewp5}
\par
\begin{center}
\rotatebox{270}{\includegraphics[width=6.0cm]{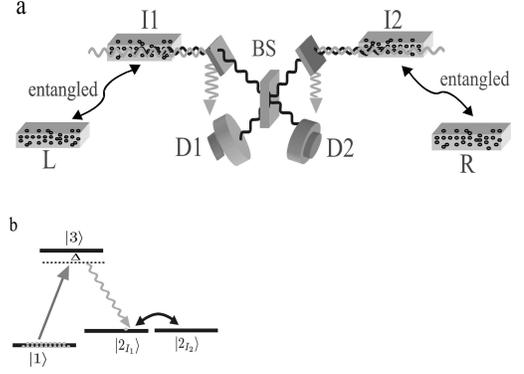}}
\end{center}
\caption{(a) Illustrative setup for the entanglement swapping. We have two
pairs of ensembles L, I$_{1}$ and I$_{2}$, R distributed at three sites L, I
and R. Each of the ensemble-pairs L, I$_{1}$ and I$_{2}$, R is prepared in
an EME state in the form of Eq. (3). The excitations in the collective modes
of the ensembles I$_{1}$ and I$_{2}$ are transferred simultaneously to the
optical excitations by the repumping pulses applied to the atomic transition
$\left| 2\right\rangle \rightarrow\left| 3\right\rangle $, and the
stimulated optical excitations, after a 50\%-50\% beam splitter, are
detected by the single-photon detectors D1 and D2. If either D1 \textit{or}
D2 clicks, the protocol is successful and an EME state in the form of Eq.
(3) is established between the ensembles L and R with a doubled
communication distance. Otherwise, the process fails, and we need to repeat
the previous entanglement generation and swapping until finally we have a
click in D1 or D2, that is, until the protocol finally succeeds. (b) The two
intermediated ensembles I$_{1}$ and I$_{2}$ can also be replaced by one
ensemble but with two metastable states I$_{1}$ and I$_{2}$ to store the two
different collective modes. The 50\%-50\% beam splitter operation can be
simply realized by a $\protect\pi/2$ pulse on the two metastable states
before the collective atomic excitations are transferred to the optical
excitations. }
\end{figure}

After successful generation of entanglement within the attenuation length,
we want to extend the quantum communication distance. This is done through
entanglement swapping with the configuration shown in Fig.~ \ref{d63reviewp5}%
. Suppose that we start with two pairs of the entangled ensembles described
by the state $\rho_{LI_{1}}\otimes\rho_{I_{2}R}$, where $\rho_{LI_{1}}$ and $%
\rho_{I_{2}R}$ are given by Eq. (\ref{21}). In the ideal case, the setup
shown in Fig.~ \ref{d63reviewp5} measures the quantities corresponding to
operators $S_{\pm}^{\dagger}S_{\pm}$ with $S_{\pm}=\left( S_{I_{1}}\pm
S_{I_{2}}\right) /\sqrt{2}$. If the measurement is successful (i.e., one of
the detectors registers one photon), we will prepare the ensembles L and R
into another EME state. The new $\varphi$-parameter is given by $\varphi
_{1}+\varphi_{2}$, where $\varphi_{1}$ and $\varphi_{2}$ denote the old $%
\varphi$-parameters for the two segment EME states. Even in the presence of
realistic noise such as the photon loss, an EME state is still created after
a detector click. The noise only influences the success probability to get a
click and the new vacuum coefficient in the EME state. The above method for
connecting entanglement can be continued to arbitrarily extend the
communication distance.

\subsubsection{ Entanglement-based communication schemes}

After an EME\ state has been established between two distant sites, we would
like to use it in the communication protocols, such as for quantum
teleportation, cryptography, or Bell inequality detection. It is not obvious
that the EME state (\ref{21}), which is entangled in the Fock basis, is
useful for these tasks since in the Fock basis it is experimentally hard to
do certain single-bit operations. In the following we will show how the EME\
states can be used to realize all these protocols with simple experimental
configurations.

\begin{figure}[ptb]
\label{d63reviewp6}
\par
\begin{center}
\includegraphics[width=6.0cm]{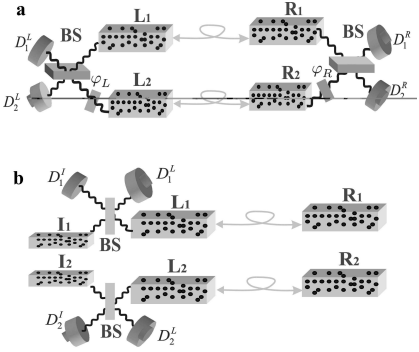}
\end{center}
\caption{(a) Schematic setup for the realization of quantum cryptography and
Bell inequality detection. Two pairs of ensembles L$_{1}$, R$_{1}$ and L$_{2}
$, R$_{2}$ have been prepared in the EME states. The collective atomic
excitations on each side are transferred to the optical excitations, which,
respectively after a relative phase shift $\protect\varphi_{L}$ or $\protect%
\varphi_{R}$ and a 50\%-50\% beam splitter, are detected by the
single-photon detectors $D_{1}^{L},D_{2}^{L}$ and $D_{1}^{R},D_{2}^{R}$. We
look at the four possible coincidences of $D_{1}^{R},D_{2}^{R}$ with $%
D_{1}^{L},D_{2}^{L}$, which are functions of the phase difference $\protect%
\varphi_{L}-\protect\varphi_{R}$. Depending on the choice of $\protect\varphi%
_{L}$ and $\protect\varphi_{R} $, this setup can realize both the quantum
cryptography and the Bell inequality detection. (b) Schematic setup for
probabilistic quantum teleportation of the atomic ``polarization'' state.
Similarly, two pairs of ensembles L$_{1}$, R$_{1}$ and L$_{2}$, R$_{2} $ are
prepared in the EME states. We want to teleport an atomic ``polarization''
state $\left( d_{0}S_{I_{1}}^{\dagger}+d_{1}S_{I_{2}}^{\dagger}\right)
\left| 0_{a}0_{a}\right\rangle _{I_{1}I_{2}}$ with unknown coefficients $%
d_{0},d_{1} $ from the left to the right side, where $S_{I_{1}}^{\dagger
},S_{I_{2}}^{\dagger}$ denote the collective atomic operators for the two
ensembles I$_{1}$ and I$_{2}$ (or two metastable states in the same
ensemble). The collective atomic excitations in the ensembles I$_{1}$, L$%
_{1} $ and I$_{2}$, L$_{2}$ are transferred to the optical excitations,
which, after a 50\%-50\% beam splitter, are detected by the single-photon
detectors $D_{1}^{I},D_{1}^{L}$ and $D_{2}^{I},D_{2}^{L}$. If there are a
click in $D_{1}^{I}$ \textit{or }$D_{1}^{L}$ and a click in $D_{2}^{I}$
\textit{or }$D_{2}^{I}$, the protocol is successful. A $\protect\pi$-phase
rotation is then performed on the collective mode of the ensemble R$_{2}$
conditional on that the two clicks appear in the detectors $D_{1}^{I}$,$%
D_{2}^{L}$ or $D_{2}^{I}$,$D_{1}^{L}$. The collective excitation in the
ensembles R$_{1}$ and R$_{2}$, if appearing, would be found in the same
``polarization'' state $\left(
d_{0}S_{R_{1}}^{\dagger}+d_{1}S_{R_{2}}^{\dagger}\right) \left|
0_{a}0_{a}\right\rangle _{R_{1}R_{2}}$. }
\end{figure}

Quantum cryptography and the Bell inequality detection are achieved with the
setup shown by Fig.~ \ref{d63reviewp6}a. The state of the two pairs of
ensembles is expressed as $\rho_{L_{1}R_{1}}\otimes\rho_{L_{2}R_{2}}$, where
$\rho_{L_{i}R_{i}}$ $\left( i=1,2\right) $ denote the same EME state with
the vacuum coefficient $c_{n}$ if we have done $n$ times entanglement
connection. The $\varphi$-parameters in $\rho_{L_{1}R_{1}}$ and $\rho
_{L_{2}R_{2}}$ are the same provided that the two states are established
over the same stationary channels. We register only the coincidences of the
two-side detectors, so the protocol is successful only if there is a click
on each side. Under this condition, the vacuum components in the EME states,
together with the state components $S_{L_{1}}^{\dagger}S_{L_{2}}^{\dagger
}\left| \text{vac}\right\rangle $ and $S_{R_{1}}^{\dagger}S_{R_{2}}^{\dagger
}\left| \text{vac}\right\rangle $, where $\left| \text{vac}\right\rangle $
denotes the ensemble state $\left| 0_{a}0_{a}0_{a}0_{a}\right\rangle
_{L_{1}R_{1}L_{2}R_{2}}$, have no contributions to the experimental results.
So, for the measurement scheme shown by Fig.~ \ref{d63reviewp4}, the
ensemble state $\rho_{L_{1}R_{1}}\otimes\rho_{L_{2}R_{2}}$ is effectively
equivalent to the following ``polarization'' maximally entangled (PME) state
(the terminology of ``polarization'' comes from an analogy to the optical
case)
\begin{equation}
\left| \Psi\right\rangle _{\text{PME}}=\left(
S_{L_{1}}^{\dagger}S_{R_{2}}^{\dagger}+S_{L_{2}}^{\dagger}S_{R_{1}}^{%
\dagger}\right) /\sqrt{2}\left| \text{vac}\right\rangle .   \label{22}
\end{equation}
The success probability for the projection from $\rho_{L_{1}R_{1}}\otimes
\rho_{L_{2}R_{2}}$ to $\left| \Psi\right\rangle _{\text{PME}}$ (i.e., the
probability to get a click on each side) is given by $p_{a}=1/[2\left(
c_{n}+1\right) ^{2}]$. One can also check that in Fig.~ \ref{d63reviewp6},
the phase shift $\psi_{\Lambda}$ $\left( \Lambda=L\text{ or }R\right) $
together with the corresponding beam splitter operation are equivalent to a
single-bit rotation in the basis $\left\{ \left| 0\right\rangle _{\Lambda
}\equiv S_{\Lambda_{1}}^{\dagger}\left| 0_{a}0_{a}\right\rangle _{\Lambda
_{1}\Lambda_{2}},\text{ }\left| 1\right\rangle _{\Lambda}\equiv
S_{\Lambda_{2}}^{\dagger}\left| 0_{a}0_{a}\right\rangle
_{\Lambda_{1}\Lambda_{2}}\right\} $ with the rotation angle $%
\theta=\psi_{\Lambda}/2$. Since we have the effective PME state and we can
perform the desired single-bit rotations in the corresponding basis, it is
clear how to use this facility to realize quantum cryptography, Bell
inequality detection, as well as teleportation (see Fig. \ref{d63reviewp6}b).

It is remarkable that all the steps of entanglement generation, connection,
and applications described above are robust to practical noise. The dominant
noise in this system is photon loss, including the contributions from the
channel attenuation, the detector and the coupling inefficiencies etc. It
the photon is lost, we will never get a click from the detectors, and we
simply repeat this failed attempt until we succeed. So this noise only
influences the efficiency to register a photon, but has no influence on the
final state fidelity if the photon is registered. Furthermore, one can show
that the nose influence on the efficiency is actually only moderate in the
sense that the required number of attempts for a successful event only
increases with the communication distance by a slow polynomial law \cite%
{Duan01b}. So we get high-fidelity quantum communication with a moderate
polynomial overhead, which is the essential advantage of the quantum
repeater protocol.

\subsubsection{Recent experimental advances}

The physics behind the above scheme for quantum repeaters is based on the
definite correlation between the forward-scattered Stokes photon and the
long-lived excitation in the collective atomic mode. The correlation comes
from the collective enhancement effect due to many-atom coherence (for a
single atom, the atomic excitation cannot be correlated with radiation in a
certain direction without the use of high-finesse cavities \cite{Duan02A}).
The entanglement generation, connection, and application schemes described
above are all based on this correlation. So the first enabling step for
demonstration of this comprehensive quantum repeater scheme is to verify
this correlation. Several exciting experiments have been reported on
demonstration of this correlation effect \cite{Kuzmich03,Lukin03,ustc03}.

The first experiment was reported from Caltech which demonstrate the
non-classical correlation between the emitted photon and the collective
atomic excitation. The collective atomic excitation is subsequently
transferred to a forward-scattered anti-Stokes photon for measurements (see
Sec. 3.2.2), so what one really detects in experiments is the correlation
between the pair of Stokes and anti-Stokes photons emitted successively. In
the Caltech experiment, the atomic ensemble is a cloud of cold atomic in a
magnetic optical trap. To experimentally confirm the correlation between the
Stokes and the anti-Stokes photons, one measures the auto-correlations $%
\tilde{g}_{1,1},\tilde{g}_{2,2}$ of the Stokes and the anti-Stokes fields
and the cross correlation $\tilde{g}_{1,2}$ between them. For any classical
optical fields (fields with well defined $P-$representations), these
correlations should satisfy the Cauthy-Schwarz inequality $\left[ \tilde{g}%
_{1,2}\right] ^{2}\leq\tilde{g}_{1,1}\tilde{g}_{2,2}$, while for
correlations between the non-classical single-photon pairs, this inequality
will be violated. In the experiment \cite{Kuzmich03}, this inequality was
measured to be strongly violated with $[\tilde{g}_{1,2}^{2}(\delta
t)=5.45\pm0.11]\nleq\lbrack \tilde{g}_{1,1}\tilde{g}_{2,2}=2.97\pm0.08]$.
Here, $\delta t$ is the time delay between the pair of \ Stokes and
anti-Stokes photons, which is $405$ ns in the initial experiment but could
be much longer (up to seconds) if one loads the atoms into a
far-off-resonant optical trap. Note that $\delta t$ is basically limited by
the spin relaxation time in the ensemble, and for implementation of quantum
repeaters it is important to get a long $\delta t$ to enable storage of
quantum information in the ensemble.

Another related experiment was reported from Harvard \cite{Lukin03}, which
uses hot atomic gas instead of the cold atomic ensemble. This experiment
also measures the correlation between the Stokes and anti-Stokes fields. The
difference is that it is not operated in the single-photon region. Instead,
both the Stokes and anti-Stokes fields may have up to thousand of photons.
In this limit, there is also some inequality need to be satisfied by the
classical fields, and the experiment measures a violation of this inequality
by about $4\%$. The other experiment with room-temperature atomic gas was
reported from USTC \cite{ustc03}, which operates in the single-photon region
as required by the quantum repeater scheme. This experiment uses a similar
detection method as the Caltech experiment, and measures a violation of the
Cauchy-Schwarz inequality with $\left[ g_{1,2}^{2}(\delta t)=4.17\pm 0.09%
\right] \nleq\left[ g_{1,1}g_{2,2}=3.12\pm0.08\right] $, where the time
delay $\delta t$ is observed to be about $2$ $\mu$s.

\subsection{Atomic ensembles for continuous variable quantum information
processing}

In continuous variable quantum information protocols, information is carried
by some observables with continuous values. There have been quite a lot of
interests in continuous variable information processing, including proposals
for continuous variable quantum teleportation, cryptography, computation,
error correction, and entanglement purification \cite%
{Braunstein+Pati-QuantInfoContVar:03}.

Here we will review some recent schemes using atomic ensembles for
realization of continuous variable quantum teleportation \cite%
{Duan002,Kuzmich00,Julsgaard01}. Note that atomic quantum teleportation (not
realized yet) typically requires strong coupling between the atom and the
photon. Collective enhancement in the atomic ensemble plays an important
role here as it significantly alleviates this stringent requirement. We will
briefly explain the idea in Ref. \cite{Duan002} which uses only coherent
light to generate continuous variable entanglement between two distant
ensembles for atomic quantum teleportation. The scheme in \cite{Duan002} has
been followed by the exciting experiment reported in Ref. \cite{Julsgaard01}
which demonstrates entanglement between two macroscopic ensembles for the
first time.

\begin{figure}[ptb]
\label{d63reviewp8}
\par
\begin{center}
\includegraphics[width=6.0cm]{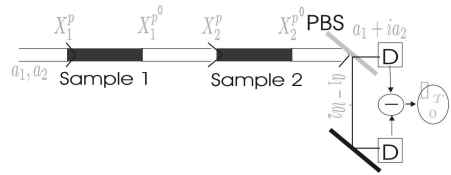}
\end{center}
\caption{Schematic setup for Bell measurements. A linearly polarized strong
laser pulse (decomposed into two circular polarization modes $a_{1},a_{2}$)
propagates successively through the two atomic samples. The two polarization
modes $\left( a_{1}+ia_{2}\right) /\protect\sqrt{2}$ and $\left(
a_{1}-ia_{2}\right) /\protect\sqrt{2}$ are then split by a polarizing beam
splitter (PBS), and finally the difference of the two photon currents
(integrated over the pulse duration $T$) is measured.}
\end{figure}

For an optical field with two circular polarization modes $a_{1}$, $a_{2}$,
one can introduce the Stokes operators by $S_{x}^{p}=\frac{1}{2}\left(
a_{1}^{\dagger}a_{2}+a_{2}^{\dagger}a_{1}\right) ,$ $S_{y}^{p}=\frac{1}{2i}%
\left( a_{1}^{\dagger}a_{2}-a_{2}^{\dagger}a_{1}\right) ,$ $S_{z}^{p}=\frac{1%
}{2}\left( a_{1}^{\dagger}a_{1}-a_{2}^{\dagger}a_{2}\right) .$ If the light
is linearly polarized along the $\overrightarrow{x}$ direction, one can
define a pair of canonical operators by $X^{p}=S_{y}^{p}/\sqrt{\left\langle
S_{x}^{p}\right\rangle },$ $P^{p}=S_{z}^{p}/\sqrt {\left\langle
S_{x}^{p}\right\rangle }$ with $\left[ X^{p},P^{p}\right] =i$. Similarly,
for a polarized atomic ensemble with the collective spin $\overrightarrow{%
S^{a}}$ pointing to the $\overrightarrow{x}$ direction, one can also define
a pair of canonical operators $X^{a}=S_{y}^{a}/\sqrt {\left\langle
S_{x}^{a}\right\rangle },$ $P^{a}=S_{z}^{a}/\sqrt{\left\langle
S_{x}^{a}\right\rangle }$ with $\left[ X^{a},P^{a}\right] =i$. When the
light passes through the atomic ensemble in an appropriate off-resonant
interaction configuration detailed in Ref. \cite{Duan002}, the continuous
variable operators defined above will transform by the following form

\begin{align}
X^{p\prime} & =X^{p}-\kappa_{c}P^{a},  \notag \\
X^{a\prime} & =X^{a}-\kappa_{c}P^{p},  \label{18} \\
P^{\beta\prime} & =P^{\beta},\text{ }\left( \beta=a,p\right) ,  \notag
\end{align}
where $\kappa_{c}$ is a parameter characterizing the interaction strength
whose typical value is around $5$.

For quantum teleportation, first one needs to generate entanglement between
two distant ensembles 1 and 2. This is done through a nonlocal Bell
measurement of the EPR operators $X_{1}^{a}-X_{2}^{a}$ and $%
P_{1}^{a}+P_{2}^{a}$ with the setup depicted by Fig.~ \ref{d63reviewp8}.
This setup measures the Stokes operator $X_{2}^{p\prime}$ of the output
light. Using Eq.~(3.5), we have $X_{2}^{p\prime}=X_{1}^{p}+\kappa_{c}\left(
P_{1}^{a}+P_{2}^{a}\right) $, so we get a collective measurement of $%
P_{1}^{a}+P_{2}^{a}$ with some inherent vacuum noise $X_{1}^{p}$. The
efficiency $1-\eta$ of this measurement is determined by the parameter $%
\kappa_{c}$ with $\eta=1/\left( 1+2\kappa_{c}^{2}\right) $. After this round
of measurements, we rotate the collective atomic spins around the $x$ axis
to get the transformations $X_{1}^{a}\rightarrow-P_{1}^{a},$ $%
P_{1}^{a}\rightarrow X_{1}^{a}$ and $X_{2}^{a}\rightarrow P_{2}^{a},$ $%
P_{2}^{a}\rightarrow -X_{2}^{a}$. The rotation of the atomic spin can be
easily obtained by applying classical laser pulses. After the rotation, the
measured observable of the first round of measurement is changed to $%
X_{1}^{a}-X_{2}^{a}$ in the new variables. We then make another round of
collective measurement of the new variable $P_{1}^{a}+P_{2}^{a}$. In this
way, both the EPR operators $X_{1}^{a}-X_{2}^{a}$ and $P_{1}^{a}+P_{2}^{a}$
are measured, and the final state of the two atomic ensembles is collapsed
into a two-mode squeezed state with variance $\delta\left(
X_{1}^{a}-X_{2}^{a}\right) ^{2}=\delta\left( P_{1}^{a}+P_{2}^{a}\right)
^{2}=e^{-2r}$, where the squeezing parameter $r$ is given by
\begin{equation}
r=\frac{1}{2}\ln\left( 1+2\kappa_{c}^{2}\right) .   \label{28}
\end{equation}
Thus, using only coherent light, we generate continuous variable
entanglement \cite{Duan004} between two nonlocal atomic ensembles. With the
interaction parameter $\kappa_{c}\approx5$, a high squeezing (and thus a
large entanglement) $r\approx2.0$ is obtainable.

To achieve quantum teleportation, first the ensembles 1 and 2 are prepared
in a continuous variable entangled state using the nonlocal Bell measurement
described above. Then, a Bell measurement with the same setup as shown by
Fig.~\ref{d63reviewp8} on the two local ensembles 1 and 3, together with a
straightforward displacement of $X_{3}^{a},$ $P_{3}^{a}$ on the sample 3,
will teleport an unknown collective spin state from the atomic ensemble 3 to
2. The teleported state on the ensemble 2 has the same form as that in the
original proposal of continuous variable teleportation using squeezing light
\cite{Braunstein981}, with the squeezing parameter $r$ replaced by Eq.~(\ref%
{28}) and with an inherent Bell detection inefficiency $\eta =1/\left(
1+2\kappa_{c}^{2}\right) $. The quality of teleportation is best described
by the fidelity, which, for a pure input state, is defined as the overlap of
the teleported state and the input state. For any coherent input state of
the sample 3, the teleportation fidelity is given by
\begin{equation}
F=1/\left( 1+\frac{1}{1+2\kappa_{c}^{2}}+\frac{1}{2\kappa_{c}^{2}}\right) .
\label{29}
\end{equation}
Equation (\ref{29}) shows that a high fidelity $F\approx96\%$ would be
possible for the teleportation of the collective atomic spin state with the
interaction parameter $\kappa_{c}\approx5$.

In the experimental demonstration \cite{Julsgaard01}, the atomic ensembles
are provided by room-temperature Cesium atomic gas in two separate glass
cells with coated wall to increase the spin relaxation time. Each cell is
about $3$ cm long, containing about $10^{12}$ atoms. The entanglement is
generated through collective Bell measurements by transmitting a coherent
light pulse as described above. To confirm and measure the generated
entanglement, one needs to transmit another verifying pulse. Trough a
homodyne detection of this verifying pulse, one can basically detect the
EPR\ variation $\Delta _{EPR}=\left[ \delta\left( X_{1}^{a}-X_{2}^{a}\right)
^{2}+\delta\left( P_{1}^{a}+P_{2}^{a}\right) ^{2}\right] /2$ \cite{Duan004},
and $\xi =1-\Delta_{EPR}$ serves as a measure of the entanglement, which is
zero for separable states and $1$ for the maximally entangled state. In this
experiment, $\xi$ is measured to be $\left( 35\pm7\right) \%$, and this
entanglement survives by about $0.5$ ms (the relaxation time is measured by
changing the time delay between the entangling and the verifying pulses).
The demonstrated entanglement will be important for the next-step
applications.

\section{Conclusions}

During the last few years the fields of atomic physics and quantum optics
have experienced an enormous progress in controlling and manipulating atoms
with lasers. This has immediate implications for quantum information
processing, since this progress allows atomic systems to fulfill the basic
requirements to implement the basic building blocks of a quantum computer.
In this article we have illustrated these statements with two particular
systems: trapped ions, neutral atoms in optical lattices and atomic
ensembles.

The physics of trapped ions is very well understood. In fact, with the
recent experimental results we can foresee no fundamental obstacle to build
a scalable quantum computer with trapped ions. Of course, technical
development may impose severe restrictions to the time scale in which this
is achieved. On the other hand, neutral atoms in optical lattices seem to be
ideal candidates to study a variety of physical phenomena by using them to
simulate other physical systems. This quantum simulation may turn out to be
the first real application of quantum information processing. Atomic
ensembles, on the other hand, are ideal to realize quantum communication
protocols (e.g. the quantum repeater, and the entanglement of distant atomic
ensembles) within setups which are considerably simpler from an experimental
point of view than the single atom and ion experiments. There are other
quantum optical systems that have experienced a very remarkable progress
during the last years, and which may equally important in the context of
quantum information. An example is cavity QED, where groups at Caltech,
Georgia Tech, Innsbruck, and Munich have trapped single atoms and ions
inside cavities, and let them interact with the cavity field, which can be
used as single (or entangled) photon(s) generators as well as to build
quantum repeaters for quantum communication.



\begin{thebibliography}{99}
\bibitem{Nielsen+Chuang-QC:00} M.~Nielsen and I.~Chuang. \newblock {\em
Quantum Computation and Quantum Information}. \newblock Cambridge University
Press, 2000.

\bibitem{Braunstein+Pati-QuantInfoContVar:03} S.~Braunstein and A.~K. Pati,
editors. \newblock {\em Quantum Information with Continuous Variables}. %
\newblock Kluwer Academic Publishers, 2003.

\bibitem{Cirac+Duan+Jaksch+Zoller-Varenna:02} J.I. Cirac, L.M. Duan,
D.~Jaksch, and P.~Zoller. \newblock Quantum optical implementation of
quantum information processing. \newblock In F.~De Martini and C.~Monroe,
editors, \emph{Proceedings of the International School of Physics ``Enrico
Fermi'' Course CXLVIII, Experimental Quantum Computation and Information}.
IOS Press, Amsterdam, 2002.

\bibitem{Levi-PhysTodayIonTrapQC:03} B.~G. Levi. \newblock {\em Physics
Today}, May 2003.

\bibitem{Cirac+Zoller-ScienceColdAtomReview:03} J.~I. Cirac and P.~Zoller. %
\newblock {\em Science}, 301:176, 2003.

\bibitem{Raimond+Brune+Haroche-collRMP:01} J.~M. Raimond, M.~Brune, and
S.~Haroche. \newblock {\em Rev. Mod. Phys.}, 73:565, 2001.

\bibitem{Lukin-AtomicEnsemblesCollRevModPhys:03} M.~D. Lukin. \newblock 2003.

\bibitem{Cirac+Zoller-QuantCompColdTrappedIons:95} J.~I. Cirac and
P.~Zoller. \newblock {\em Phys. Rev. Lett.}, 74(20):4091, 1995.

\bibitem{Steane-ReviewIonTrap:97} A.~Steane. \newblock {\em Appl. Phys. B},
64:623, 1997.

\bibitem{Schmidt-Kaler+BlattETAL-CiracZollerGate:03} F.~Schmidt-Kaler, H.~H{%
\"a}ffner, M.~Riebe, S.~Gulde, G.P.T. Lancaster, T.~Deuschle, C.~Becher,
C.F. Roos, J.~Eschner, and R.~Blatt. \newblock {\em Nature}, 422:408, 2003.

\bibitem{Leibfried+DeMarcoETAL-ExpDemTwoIonPhaseGate:03} D.~Leibfried,
B.~DeMarco, V.~Meyer, D.~Lucas, M.~Barrett, J.~Britton, W.~M. Itano,
B.~Jelenkovi, C.~Langer, T.~Rosenband, and D.~J. Wineland. \newblock {\em
Nature}, 422:412, 2003.

\bibitem{Leibfried+DeMarcoETAL-NISTreview:03} D.~Leibfried, B.~DeMarco,
V.~Meyer, M.~Rowe, A.~Ben-Kish, M.~Barrett, J.~Britton, J.~Hughes, W.~M.
Itano, B.~M. Jelenkovic, C.~Langer, D.~Lucas, T.~Rosenband, and D.~J.
Wineland. \newblock {\em J. Phys. B}, 36:599, 2003.

\bibitem{Garcia-Ripoll+Zoller+Cirac-SpeedOptimizedIonGate:03} J.~J. Garc%
\'{\i}a-Ripoll, P.~Zoller, and J.~I. Cirac. \newblock {\em Phys. Rev. Lett.}%
, 91:157901, 2003.

\bibitem{Duan01} L.-M. Duan, J.~I. Cirac, and P.~Zoller. \newblock {\em
Science}, 292:1695, 2001.

\bibitem{Gardiner+Zoller-QuantumNoise:99} C.~W. Gardiner and P.~Zoller.
\newblock {\em Quantum Noise: A Handbook of Markovian and Non-Markovian Quantum
Stochastic Methods with Applications to Quantum Optics}. \newblock Springer,
1999.

\bibitem{Wineland+MonroeETAL-NISTbible:98} D.~J. Wineland, C.~Monroe, W.~M.
Itano, D.~Leibfried, B.~E. King, and D.~M. Meekhof. \newblock {\em RES J.
NIST}, 103:259, 1998.

\bibitem{Cirac+Zoller-ScalableQCTrappedIons:00} J.~I. Cirac and P.~Zoller. %
\newblock {\em Nature}, 404:579, 2000.

\bibitem{Kielpinksi+Monroe+Wineland-ScalableQC:02} D.~Kielpinksi, C.~Monroe,
and D.~J. Wineland. \newblock {\em Nature}, 417:709, 2002.

\bibitem{Steane+RoosETAL-Speeion-quanproc:00} A.~Steane, C.~F. Roos,
D.~Stevens, A.~Mundt, D.~Leibfried, F.~Schmidt-Kaler, and R.~Blatt. %
\newblock {\em Phys. Rev. A}, page 042305, 2000.

\bibitem{Sorensen+Molmer-EntanglementQCIonsThermalMotion:00} A.~S{\o }rensen
and K.~M{\o }lmer. \newblock {\em Phys. Rev. A}, 62:022311, 2000.

\bibitem{Sorensen+Molmer-PRLQCIonThermalMotion:99} A.~S\o rensen and K.~M\o %
lmer. \newblock {\em Phys. Rev. Lett.}, 82(9):1971, 1999.

\bibitem{Knight00} D.~Jonathan, M.~B. Plenio, and P.~L. Knight. \newblock
{\em Phys. Rev. A}, 62:042307, 2000.

\bibitem{Milburn+SchneiderETAL-trapquancompwith:00} G.~J. Milburn,
S.~Schneider, and D.~F.~V. James. \newblock {\em Fortschritte der Physik},
48:801, 2000.

\bibitem{DeMarco+Ben-KishETAL-Expedemocontwave:02} B.~DeMarco, A.~Ben-Kish,
D.~Leibfried, V.~Meyer, M.~Rowe, B.~M. Jelenkovic, W.~M. Itano, J.~Britton,
C.~Langer, T.~Rosenband, and D.~J. Wineland. \newblock {\em Phys. Rev. Lett.}%
, 89:267901--1, 2002.

\bibitem{Jaksch+BruderETAL-ColdBosonicAtomOpticalLattices:98} D.~Jaksch,
C.~Bruder, J.I. Cirac, C.~Gardiner, and P.~Zoller. \newblock {\em Phys. Rev.
Lett.}, 81(15):3108, 1998.

\bibitem{JakschETAL-EntanglementColdCollision:99} D.~Jaksch, H.J. Briegel,
J.I. Cirac, C.W. Gardiner, and P.~Zoller. \newblock {\em Phys. Rev. Lett.},
82:1975, 1999.

\bibitem{Bloch-SF-Mott:02} M.~Greiner, O.~Mandel, T.~Esslinger, T.W. H{\"a}%
nsch, and I.~Bloch. \newblock {\em Nature}, 415:39, 2002.

\bibitem{Bloch-CCEntanglement:03} O.~Mandel, M.~Greiner, A.~Widera, T.~Rom,
T.W. H{\"a}nsch, and I.~Bloch. \newblock {\em Nature}, 425:937, 2003.

\bibitem{Brennen+Caves-Deutsch-QuantumLogicGateOpticalLattice:99} %
G.~Brennen, C.~Caves, P.~Jessen, and I.~Deutsch. \newblock {\em Phys. Rev.
Lett.}, 82:1060, 1999.

\bibitem{Brennen+Deutsch-Williams-QuanLogiTrapAtom:02} G.K. Brennen, I.H.
Deutsch, and C.J. Williams. \newblock {\em Phys. Rev. A}, 65:022313, 2002.

\bibitem{JakschETAL-RydbergGate:00} D.~Jaksch, J.~I. Cirac, P.~Zoller, S.L.
Rolston, R.~Cote, and M.~D. Lukin. \newblock {\em Phys. Rev. Lett.},
85:2208, 2000.

\bibitem{Sachdev-QuantPhaseTransition:99} S.~Sachdev. \newblock {\em Quantum
Phase Transitions}. \newblock Cambridge University Press, Cambridge, 1999.

\bibitem{CalarcoETAL-QuantumGateCollisions:00} T.~Calarco, E.A. Hinds,
D.~Jaksch, J.~Schmiedmayer, J.I. Cirac, and P.Zoller. \newblock {\em Phys.
Rev. A}, 61:22304, 2000.

\bibitem{Schlosser+Grangier-TwoAtomTrap:01} N.~Schlosser, G.~Reymond,
I.~Protsenko, and P.~Grangier. \newblock {\em Nature}, 411:1024, 2001.

\bibitem{Buchkremer+Dumke+Birkl+Ertmer-QIMicrofabricatedOpticalElements:02} %
F.B.J. Buchkremer, R.~Dumke, M.~Volk, T.~Muether, G.~Birkl, and W.~Ertmer. %
\newblock {\em Laser Physics}, 12:736, 2002.

\bibitem{Raussendorf+Briegel-OneWayQuantumComputer:01} R.~Raussendorf and
H.~J. Briegel. \newblock {\em Phys. Rev. Lett.}, 86(22):5188, 2001.

\bibitem{Jane+VidalETAL-Simulation:03} E.~Jane, G.~Vidal, W.~D{\"u}r, and
P.~Zoller. \newblock {\em Quantum Information and Computation}, 1:15, 2003.

\bibitem{Lloyd-UniversalQuantumSimulators:96} S.~Lloyd. \newblock {\em
Science}, 273:1073, 1996.

\bibitem{Lukin002} M.~D. Lukin, M.~Fleischhuaer, R.~Cote, L.-M. Duan,
D.~Jaksch, J.~I. Cirac, and P.~Zoller. \newblock {\em Phys. Rev. Lett.},
87:037901, 2001.

\bibitem{Duan01b} L.-M. Duan, M.~D. Lukin, J.~I. Cirac, and P.~Zoller. %
\newblock {\em Nature}, 414:413, 2001.

\bibitem{Kuzmich00} A.~Kuzmich and E.~S. Polzik. \newblock {\em Phys. Rev.
Lett.}, 85(26):5639, 2000.

\bibitem{Duan002} L.-M. Duan, J.~I. Cirac, P.~Zoller, and E.~S. Polzik. %
\newblock {\em Phys. Rev. Lett.}, 85(26):5643, 2000.

\bibitem{FL00} M.~Fleischhauer and M.~D. Lukin. \newblock {\em Phys. Rev.
Lett.}, 84(22):5094, 2000.

\bibitem{Julsgaard01} B.~Julsgaard, A.~Kozhekin, and E.~S. Polzik. \newblock
{\em Nature}, 413:400, 2001.

\bibitem{Duan02} L.-M Duan. \newblock {\em Phys. Rev. Lett.}, 88:170402,
2002.

\bibitem{Kuzmich03} A.~Kuzmich, W.~P. Bowen, A.~D. Boozer, A.~Boca, C.~W.
Chou, L.-M. Duan, and H.~J. Kimble. \newblock {\em Nature}, 423:731, 2003.

\bibitem{Lukin03} C.~H. van~der Wal, M.~D. Eisaman, A.~Andre, R.~L.
Walsworth, D.~F. Phillips, A.~S. Zibrov, and M.~D. Lukin. \newblock {\em
Science}, 301:196, 2003.

\bibitem{ustc03} W.~Jiang, C.~Han, P.~Xue, L.~M. Duan, and G.~C. Guo. %
\newblock {\em quant-ph/0309175}.

\bibitem{Hald99} J.~Hald, J.~L. Sorensen, C.~Schori, and E.S. Polzik. %
\newblock {\em Phys. Rev. Lett.}, 83:1319, 1999.

\bibitem{Hau99} L.~V. Hau, S.~E. Harris, Z.~Dutton, and C.~H. Behroozi. %
\newblock {\em Nature}, 397:594, 1999.

\bibitem{Scully99} M.~M. Kash, V.~A. Sautenkov, A.~S. Zibrov, L.~Hollberg,
G.~R. Welch, M.~D. Lukin, Y.~Rostovtsev, E.~S. Fry, and M.O. Scully. %
\newblock {\em Phys. Rev. Lett.}, 82:5229, 1999.

\bibitem{Kuzmich001} A.~Kuzmich, L.~Mandel, and N.~P. Bigelow. \newblock
{\em Phys. Rev. Lett.}, 85(8):1594, 2000.

\bibitem{Briegel98} H.~J. Briegel, W.~Dur, J.~I. Cirac, and P.~Zoller. %
\newblock {\em Phys. Rev. Lett.}, 81(26):5932, 1998.

\bibitem{Ekert91} A.~Ekert. \newblock {\em Phys. Rev. Lett.}, 67(6):661,
1991.

\bibitem{Bennett93} C.~H. Bennett, G.~Brassard, C.~Crepeau, R.~Jozsa,
A.~Peres, and W.~K. Woothers. \newblock {\em Phys. Rev. Lett.}, 70(13):1895,
1993.

\bibitem{Bennett96} C.~H. Bennett, D.~P. Divincenzo, J.~A. Smolin, and W.~K.
Wootters. \newblock {\em Phys. Rev. A}, 54(5):3824, 1996.

\bibitem{Preskill} J.~Preskill. \newblock {\em
http://www.theory.caltech.edu/people/preskill/ph229/}, 2001.

\bibitem{Cirac97} J.~I. Cirac, P.~Zoller, H.~J. Kimble, and H.~Mabuchi. %
\newblock {\em Phys. Rev. Lett.}, 78(16):3221, 1997.

\bibitem{Enk98} S.J.~Van Enk, J.I. Cirac, and P.~Zoller. \newblock {\em
Science}, 279:205, 1998.

\bibitem{Ye99} J.~Ye, D.~W. Vernooy, and H.J. Kimble. \newblock {\em Phys.
Rev. Lett.}, 83:4987, 1999.

\bibitem{Rempe02} A.~Kuhn, M.~Hennrich, and G.~Rempe. \newblock {\em Phys.
Rev. Lett.}, 89:067901, 2002.

\bibitem{Chapman03} J.~A. Sauer, K.~M. Fortier, M.~S. Chang, C.~D. Hamley,
and M.~S. Chapman. \newblock {\em quant-ph/0309052}.

\bibitem{Kimble03} J.~McKeever, A.~Boca, A.~D. Boozer, J.~R. Buck, and H.~J.
Kimble. \newblock {\em Nature}, 425:268, 2003.

\bibitem{Cab99} C.~Cabrillo, J.~I. Cirac, P.~G-Fernandez, and P.~Zoller. %
\newblock {\em Phys. Rev. A}, 59(2):1025, 1999.

\bibitem{Duan02A} L.-M. Duan, J.I. Cirac, and P.~Zoller. \newblock {\em
Phys. Rev. A}, 66:023818, 2002.

\bibitem{Duan004} L.-M. Duan, G.~Giedke, J.~I. Cirac, and P.~Zoller. %
\newblock {\em Phys. Rev. Lett.}, 84:2722, 2000.

\bibitem{Braunstein981} S.~L. Braunstein and H.~J. Kimble. \newblock {\em
Phys. Rev. Lett.}, 80(4):869, 1998.
\end{thebibliography}



\end{document}